\documentclass[%
 reprint,
 amsmath,amssymb,
 aps,
]{revtex4-2}

\usepackage{graphicx}
\usepackage[caption = false]{subfig}
\usepackage{dcolumn}
\usepackage{bm}
\usepackage{xcolor}

\usepackage[normalem]{ulem}  
\usepackage{xcolor}          

\begin{document}

\preprint{APS/123-QED}

\title{Probing Multipolar Order in the Candidate Altermagnet MnF$_2$ through the Elastocaloric Effect under Strain}

\author{Rahel Ohlendorf$^{1,2}$}\email{Rahel.Ohlendorf@cpfs.mpg.de}
\author{Luca Buiarelli$^3$}
\author{Hilary M. L. Noad$^1$}
\author{Andrew P. Mackenzie$^{1,4}$}
\author{Rafael M. Fernandes$^{5,6}$}
\author{Turan Birol$^3$}
\author{Jörg Schmalian$^{7,8}$}
\author{Elena Gati$^{1,2,9}$}\email{e.gati@physik.uni-frankfurt.de}

\address{$^{1}$ Max Planck Institute for Chemical Physics of Solids, 01187 Dresden, Germany}
\address{$^{2}$ Institut für Festkörper- und Materialphysik, Technische Universität Dresden, 01062 Dresden, Germany}
\address{$^{3}$ Department of Chemical Engineering and Materials Science,
University of Minnesota, Minneapolis, Minnesota 55455, USA}
\address{$^{4}$ Scottish Universities Physics Alliance, School of Physics and Astronomy, University of St Andrews, St Andrews KY16 9SS, UK}
\address{$^{5}$ Department of Physics, The Grainger College of Engineering,
University of Illinois Urbana-Champaign, Urbana, IL 61801, USA}
\address{$^{6}$ Anthony J. Leggett Institute for Condensed Matter Theory, The Grainger College of Engineering,
University of Illinois Urbana-Champaign, Urbana, IL 61801, USA}
\address{$^{7}$ Institute for Theory of Condensed Matter, Karlsruhe Institute of Technology, 76131 Karlsruhe, Germany}
\address{$^{8}$ Institute for Quantum Materials and Technologies,
Karlsruhe Institute of Technology, 76126 Karlsruhe, Germany}
\address{$^{9}$ Institute of Physics, Goethe University, 60438 Frankfurt/Main, Germany}

\date{\today}

\begin{abstract}
Altermagnets break a combination of time-reversal and rotational symmetries without generating a net magnetization. As such, the order parameter of $d$-wave altermagnets has the same symmetry as magnetic multipoles, and couples to the product of a magnetic field and uniaxial strain. We combine elastocaloric experiments, free-energy modeling, and first-principles calculations on MnF$_2$ to establish a thermodynamic probe of the predicted finite-temperature altermagnetic critical point. These results pave the way to explore altermagnetic quantum criticality in $d$-wave materials and beyond.
\end{abstract}

\maketitle

\textit{Introduction - }Symmetry analysis is a cornerstone of condensed-matter physics, providing a rigorous framework for identifying broken-symmetry states. In the context of collinear magnetism, a recent classification based on spin groups has revealed a rich variety of magnetic orders that extend far beyond conventional dipolar ferro- and antiferromagnets. Among these, so-called altermagnets (AMs) \cite{Smejkal22b,Smejkal22,Jungwirth25} have garnered significant attention recently due to their unique electronic structure, characterized by an alternating spin polarization \cite{Smejkal2020_AM,Hayami19} of the energy bands combined in certain cases with zero net magnetization \cite{Fernandes2024_AM}. This property opens pathways toward novel functionalities in metallic spintronic devices \cite{Smejkal22,Jungwirth2025spintronics}. 

The distinctive features of AMs arise from their compensated magnetic structures, which preserve combinations of time-reversal and rotational point-group symmetries. Because the relevant symmetry constraints involve operations both in spin space and in real space \cite{Brinkmann66,Litvin74}, AM order parameters can be interpreted in terms of higher-rank multipole orders \cite{Bho24,Fernandes2024_AM,McC24,Schiff2024,Jaeschke2025,Buiarelli2025}, instead of a conventional dipole order \cite{Hayami20}.

The symmetries of an ordered phase impose strict constraints on its allowed response functions. While recent studies have focused on spectroscopic and transport signatures of the unconventional symmetry breaking in AMs \cite{Jungwirth25b,Hariki24,Krempasky24}, thermodynamic characterization remains less explored. However, such measurements offer fundamental insights, since the application of a conjugate field $\hat{h}$ — an external perturbation that transforms like the order parameter — causes a distinct thermodynamic response that is a direct consequence of the broken symmetry in the ordered state. One key prediction is that a sharp phase transition at $\hat{h}=0$ and temperature $T_c$ becomes a crossover at finite $\hat{h}$, with the associated crossover temperature $T^*$ obeying the scaling relation
\begin{equation}
T^*-T_c\,\propto\,\hat{h}^{1/(\beta \delta)},
\label{eq:Tstar}
\end{equation}
where $\beta$ and $\delta$ are the critical exponents of the transition that determine, as usual,  the change of the order parameter with temperature and conjugate field, respectively.  Moreover, near the critical point at ($\hat{h}=0$, $T_c$), the entropy is expected to peak  and to decrease as one applies the external conjugate field $\hat{h}$. 

With the exception of ferromagnets and nematics, where $\hat{h}$ corresponds to a uniform magnetic field and uniaxial strain, the physical realization of a conjugate field is often challenging, even if the magnetic order is collinear. For instance, in a two-sublattice antiferromagnet that breaks translational symmetry, $\hat{h}$
is a staggered magnetic field that alternates between the sublattices. A distinctive feature of AMs with a $d$-wave component (or, equivalently, a ferroic order of emergent magnetic octupole moments) is that piezomagnetism \cite{Moriya59,Dzyaloshinskii57,Tavger58} allows $\hat{h}$ to be realized through a combination of magnetic field and strain \cite{Steward23,McC24,Li2024strain,Yershov24,Khodas25,Chakarborty2025}, enabling thermodynamic measurements to reveal the nature of the broken symmetry. 

\begin{figure}
\includegraphics[width=0.9\columnwidth]{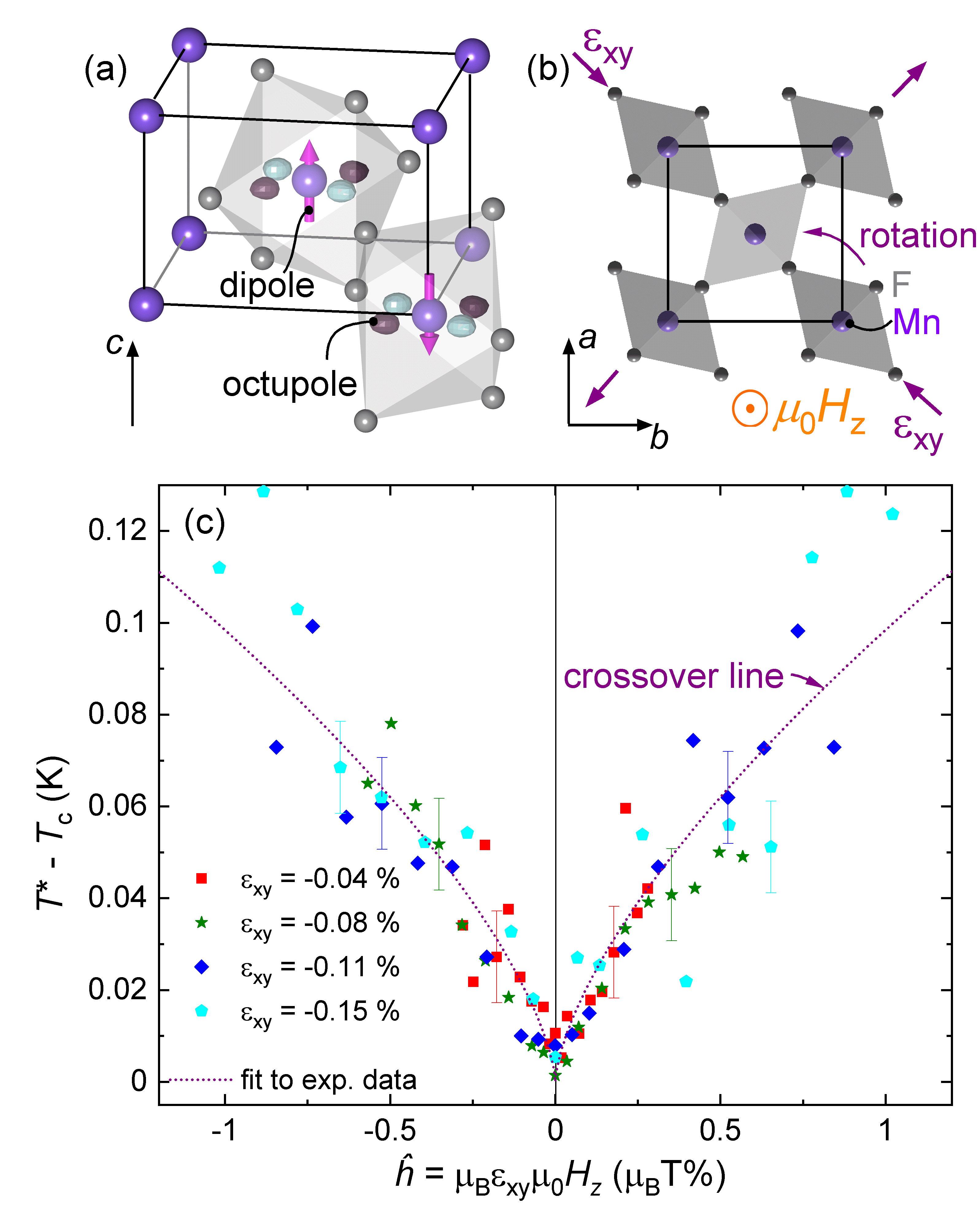}
 \caption{\label{fig:1} (a) Emergent ferro-octupolar order \cite{Bho24,Buiarelli2025} of the AM candidate MnF$_2$. The pink arrows depict the antiferromagnetic ordering of dipoles, whereas the ferroically ordered octupoles are depicted in purple and cyan. (b) A top view of the crystal structure. The F ions (gray circles) create an octahedral environment around the Mn atoms (purple circles); as a result, the two Mn sublattice sites are related by a nonsymmorphic symmetry involving a $90^\circ$ rotation and a half-translation. A strain $\varepsilon_{xy}$ will break this rotational symmetry. The conjugate field to the AM order, $\hat{h}$, is composed of $\varepsilon_{xy}$ and a magnetic field along the $c$~axis, $\mu_0 H_z$. (c) Experimentally determined crossover lines $T^*-T_c$ of MnF$_2$ as a function of conjugate field $\hat{h}=\mu_B\mu_0 H_z \varepsilon_{xy}$. The experimental data, extracted from the data in Fig.\,\ref{fig:2} and in End Matter, follow the expectation for an AM critical point at $\mu_B\mu_0 H_z \varepsilon_{xy}\,=\,0$. The data are well described by a fit to Eq.~\eqref{eq:crossover-expanded} with the mean-field exponent $1/(\beta \delta)\,=\,2/3$. For clarity, error bars are shown only for representative data points.}
\end{figure}

In this work, we experimentally verify such key thermodynamic predictions resulting from the symmetry breaking in a $d$-wave AM. This type of AM is proposed to exist in rutile insulating compounds \cite{Smejkal22,Roig24,Bho24,Morano24,Antonenko2025}, such as MnF$_2$ and CoF$_2$, which have $d_{xy}$-wave symmetry,  and metallic systems \cite{Jiang25,Sun25}
. Based on the symmetry analysis, the rutile systems exhibit an emergent ferroic time-reversal symmetry breaking octupolar order \cite{Bho24} (see Figs.~\ref{fig:1}\,(a) and (b)).  These  symmetry considerations allow us to  identify the relevant conjugate field $\hat{h}$ as a combination of a magnetic field along the $z$~axis, $\mu_0 H_z$, and a symmetry-breaking shear strain $\varepsilon_{xy}$, i.e.,
\begin{equation}
\hat{h}=\mu_B\varepsilon_{xy} \mu_0H_z.
    \label{eq:conj_fieldAM}
\end{equation}



As a method of choice to explore the thermodynamics of AMs \cite{Patri19,McC24,Yershov24} experimentally, we focus here on the elastocaloric effect (ECE) -- the adiabatic temperature change induced by strain: 

\begin{equation}
    \eta_{ij} = \left(\frac{\partial T}{\partial \varepsilon_{ij}}\right)_S=-\frac{T}{C_{\varepsilon_{ij}}}\left(\frac{\partial S}{\partial \varepsilon_{ij}}\right)_T,
\end{equation}
which is thermodynamically related to the temperature ($T$) and strain ($\varepsilon_{ij}$) derivative of the entropy ($S$).  $C_{\varepsilon_{ij}}\,=\,-T\left(\frac{\partial S}{\partial T}\right)_{\varepsilon_{ij}}$ is the specific heat at constant strain. The ECE has recently emerged as a highly sensitive probe of symmetry breaking in quantum materials, enabled by recent advances in \textit{in situ}-tunable uniaxial pressure cells \cite{Hicks14,Ike18,Li22,Ye22,Gati23,Ye24,Liu24,Lieberich25}. Despite theoretical results showing the connection between ECE and the AM susceptibility \cite{Chakarborty2025,Steward25}, the ECE has not yet been experimentally explored in the context of the multipolar order of AMs. 

Here, we establish the sensitivity of the ECE to AM symmetries through experiments on MnF$_2$ combined with symmetry analysis, free energy modeling and density functional theory (DFT). MnF$_2$ is particularly well suited for our study owing to its well-characterized magnetic order below the accessible transition temperature $T_c^{(0)}\,=\,67.5\,\mathrm{K}$~\cite{Erickson53,Morano24,Schweika02}, its known piezomagnetism \cite{Bor60,Bar88} reflecting the underlying spatial and spin symmetries, prior evidence of aspherical magnetization density around Mn ions \cite{Costa89}, and the recent observation of AM magnon splitting \cite{Faure25}.

Our key result is the experimental identification of the predicted crossover lines in MnF$_2$ [see Fig.~\ref{fig:1}\,(c)]. We complement this by identifying signatures of entropy accumulation in the vicinity of the transition temperature. Furthermore, we show through DFT calculations that the observed thermodynamic response can be microscopically explained by a weakly spin–orbit-coupled AM insulator with a small degree of off-stoichiometry. Our results provide a road map for future thermodynamic investigations of strain-tuned AMs, in particular for experimental explorations of AM quantum criticality \cite{Steward25,Chakarborty2025} in both insulators and metals.

\begin{figure}
\includegraphics[width=\columnwidth]{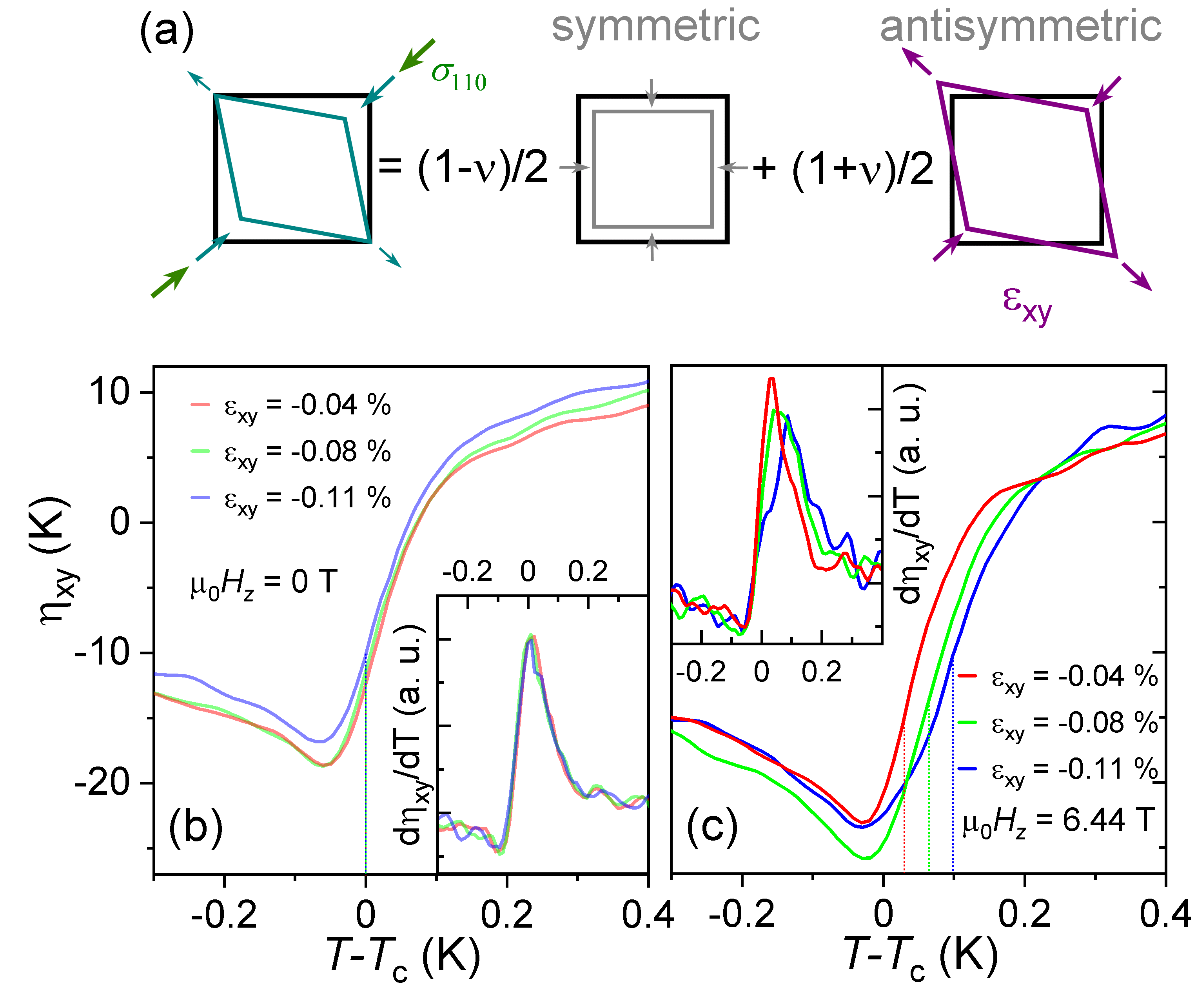}
\caption{\label{fig:2}  (a) The application of a uniaxial stress along the [110] direction, $\sigma_{110}$, leads to an induced strain that can be described by the superposition of a symmetric strain and an antisymmetric strain. The latter is denoted by $\varepsilon_{xy}$, and $\nu$ is the Poisson ratio. (b),(c) ECE data $\eta_{xy}$ on MnF$_2$ as a function of the relative temperature $T-T_c$. $T_c$ is the transition temperature including non-AM shifts in strain and field (see the text). While for $\mu_0 H_z\,=\,$ 0\,T (b) the ECE features occur at the same temperature (see dotted lines, representing the point of steepest slope), they move to higher temperature with higher compression for 6.44\,T (c), as expected for a system with AM order. The insets in (b) and (c) show the temperature derivatives of the ECE data in the respective plots. The maxima of the derivatives mark the points of steepest slope in the ECE data.}
\end{figure}

\textit{Results and discussion - } In order to induce a $\varepsilon_{xy}$ strain experimentally, we apply a uniaxial stress along the crystallographic [110] direction, $\sigma_{110}$ (see Fig.~\ref{fig:2}\,(a)). Such a stress generates both the symmetry-breaking $\varepsilon_{xy}$ component and a fully symmetric component, $1/2\left(\varepsilon_{xx}+\varepsilon_{yy}\right)$. Importantly, for any generic symmetry of the order parameter, the symmetric strain can always shift $T_c^{(0)}$ linearly, as can $(\mu_0 H_z)^2$ (see Supplemental Material~\cite{SI26}\nocite{Boo76,Ike78,Torrent2008Apr,Marques2012Oct,Gonze2016Aug,Gonze2020Mar,Romero2020Mar,Jollet2014Apr,Amadon2008,Verstraete2001,Noa23,Jerzembeck22,Mel70,Ike19,Sto11,Straquadine20,Jerzembeck24,Har72}, Sec.\,\ref{sec:strain-decomposition}). Thus, the extended AM free energy, including all terms relevant for experiments under $\sigma_{110}$, reads as~\cite{Steward23,Steward25,Chakarborty2025}

\begin{equation}
f=\frac{k_{B}}{2}\left(T-T_c\right)\Phi^{2}+\frac{u}{4}\Phi^{4}-\lambda\hat{h}\Phi
     \label{eq:free-energy-expanded}
\end{equation}
with $T_{c}=T_c^{(0)}+a_0 \frac{1}{2}(\varepsilon_{xx}+\varepsilon_{yy})+a_2 (\mu_0H_z)^2$.
$T_{c}^{\left(0\right)}$ is the transition temperature at zero strain and magnetic field, and $\Phi$ is the AM order parameter, which couples to its conjugate field with a coupling constant $\lambda$ (note that per definition $\lambda$ is dimensionless here).
Since $(\varepsilon_{xx}+\varepsilon_{yy})  \propto \sigma_{110}  \propto \varepsilon_{xy}$ (see Supplemental Material, Sec.\,\ref{sec:strain-decomposition}), we can analyze our results in terms of any of these three variables, with the understanding that the physical external control parameter is $\sigma_{110}$. Since our primary interest is on the altermagnetic term, hereafter we choose to express our results in terms of $\varepsilon_{xy}$. Consequently, when $T_c$ is plotted as a function of $\varepsilon_{xy}$, it also has a linear contribution, resulting in the following equation for $T_c$: $T_{c}=T_{c}^{\left(0\right)}+a_{1}\varepsilon_{xy}+a_{2}(\mu_{0}H_{z})^{2}$.
We obtain (see Supplemental Material, Sec.\,\ref{sec:crossover-theory})
\begin{equation}
    T^{*}=T_{c}+T_{c}\left(\frac{\lambda\hat{h}}{k_B T_{c}}\right)^{1/(\beta\delta)}
    \label{eq:crossover-expanded}
\end{equation}
which is Eq.~\eqref{eq:Tstar} given above.


Figures \ref{fig:2}\,(b) and (c) show thermodynamic ECE data $\eta_{xy}$, which reveal distinct features that we trace as a function of relative temperature $T-T_c$, for $\mu_0 H_z=0\,{\rm T}$ (b) and $6.44\,\rm{T}$ (c) [using $T_c^{(0)}=67.467(3)\,\rm{K}$, $a_2=-0.01585(14)\,\mathrm{K/T^2}$, $a_1=-59(3)\,\mathrm{K}$, determined from a global fit; see Supplemental Material Sec.\,\ref{sec:globalfit}]. The characteristic temperatures $T_c$ (at $\hat{h}=0$) and $T^\star$ (at $\hat{h}\neq 0$) are defined by the point of steepest slope in $\eta_{xy}(T)$, which corresponds to a maximum in the derivative of $\eta_{xy}(T)$ (see insets). In each panel, we show $\eta_{xy}(T-T_c)$ at three different compressive $\varepsilon_{xy}$ values. In zero field, the steplike feature appears at $T=T_c$ for all $\varepsilon_{xy}$. Under a fixed finite field, however, the feature shifts to higher $T$ with increasing $\varepsilon_{xy}$ [dotted lines in Fig.\,\ref{fig:2}\,(c)], indicating a combined action of $\varepsilon_{xy}$ and $\mu_0 H_z$.



The result of $T^\star-T_c$ as a function of $\hat{h}$, compiled from all datasets that are shown in Figs.\,\ref{fig:eps_data} and \ref{fig:H_data} in End Matter, is presented in Fig.~\ref{fig:1}\,(c). Importantly, the data exhibit a sharp kink at $\hat{h}\,=\,0$, are symmetric around that point, and collapse for all different $\varepsilon_{xy}$ onto a single curve within the experimental resolution. This behavior matches the expectation for the crossover lines from the analysis of the AM free energy. Fitting the data in Fig.~\ref{fig:1}(c) with Eq.~\eqref{eq:crossover-expanded} using  the mean-field critical exponent of $1/(\beta \delta)=2/3$ yields $\lambda\,=\,0.56(5)$ [while a fit using the 3D Ising critical exponent of $1/(\beta \delta)\approx0.64$ yields $\lambda\,=\,0.35(5)$; see Supplemental Material Sec.\,\ref{sec:crossover-theory}].

\begin{figure}
\includegraphics[width=\columnwidth]{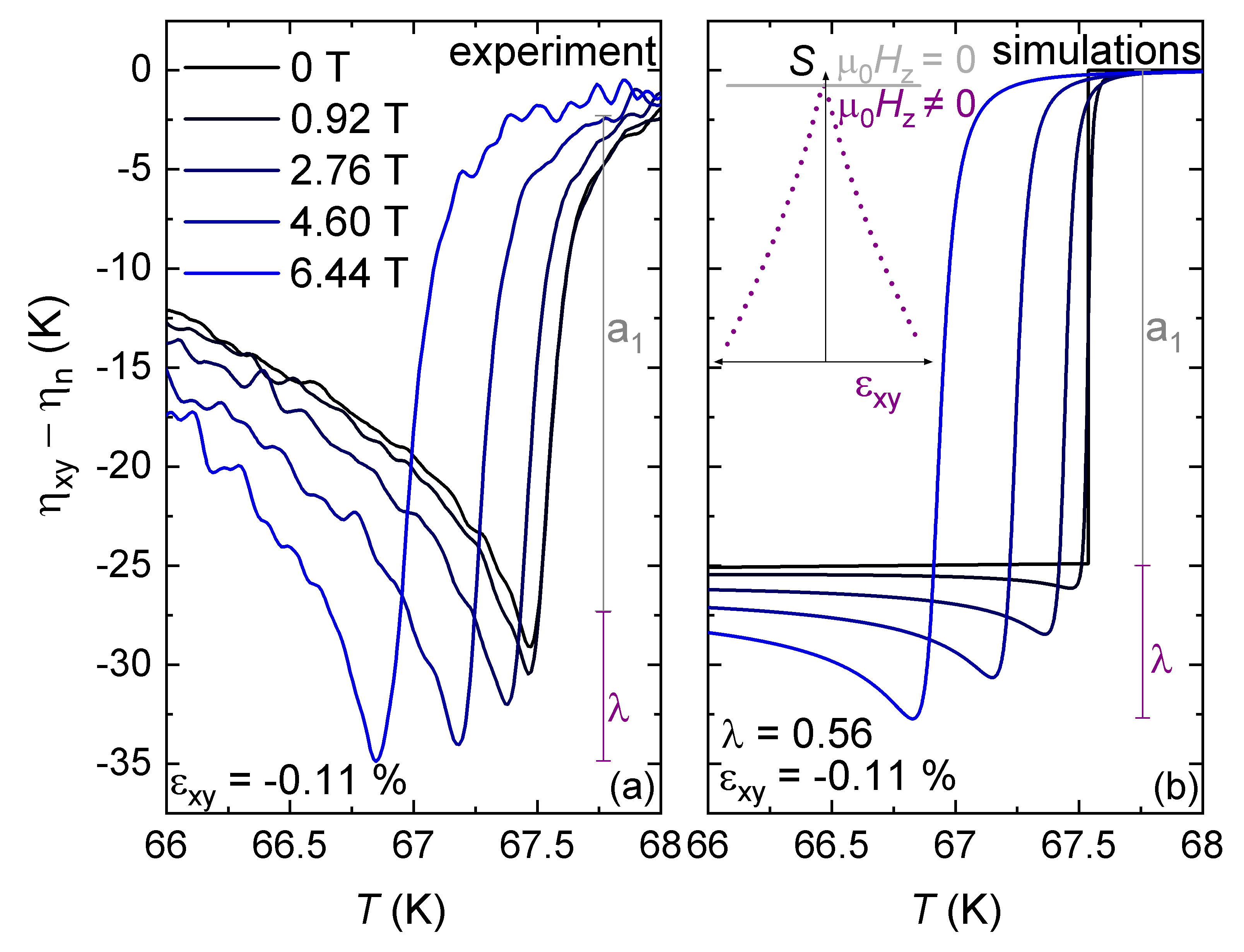}
\caption{\label{fig:3}  Comparison of (a) experimental $\eta_{xy}-\eta_n$ (with $\eta_n$ the high-temperature background; see End Matter) of MnF$_2$ as a function of $T$ at a fixed strain of $\varepsilon_{xy}\,=\,-0.11\,\%$ and different fields up to 6.44\,T  with (b) mean-field simulations for $\eta_{xy}$, based on the free energy Eq.~\eqref{eq:free-energy-expanded}. The contributions to $\eta_{xy}$ due to $a_1$ and $\lambda$ are marked by vertical bars. The AM contribution associated with $\lambda$ leads to a negative change in $\eta_{xy}$ under finite fields, as the entropy $S$ grows toward a maximum at zero strain (see the inset).}
\end{figure}

We further support the thermodynamic evidence for AM symmetries of $\Phi$ by identifying AM signatures directly in the magnitude of $\eta_{xy}$. In Fig.~\ref{fig:3}, we compare the experimental $\eta_{xy}(T)$ at fixed strain $\varepsilon_{xy}=-0.11\,\%$ and varying $\mu_0H_z$ with simulations based on the mean-field free energy Eq.~\eqref{eq:free-energy-expanded}, with $a_1, a_2$, and $\lambda$ as above. Two distinct contributions to $\eta_{xy}(T)$ can be identified. First, in zero field, both experiment and theory show a jumplike change at $T_c$, consistent with the Ehrenfest relation $\Delta \eta_{xy}\propto$ d$T_c/$d$\varepsilon_{xy}=a_1$ [see Eq.~\eqref{eq:Ehrenfest}], i.e., a field-independent, non-AM contribution to $\eta_{xy}(T)$, since it does not depend on the AM coupling $\lambda$. Second, for finite fields, we find that both experiment and simulations show a stronger negative response upon approaching $T^*$ from below (note that the general suppression of $T^*$ is a consequence of the $a_2$ parameter). On a quantitative level, there is a good agreement between simulations and experiment: The minimum of $\eta_{xy}$ decreases in both by $\approx 6$–$8$\,K when $\mu_0H_z$ increases from 0 to 6.44 T.

This change in ECE magnitude can be attributed to the AM term in the free energy, i.e., is a consequence of $\lambda$. The change in $\eta_{xy}$ with $\mu_0 H_z$ at fixed $\varepsilon_{xy}$ directly reflects the entropy accumulation at the critical point ($\varepsilon_{xy} \mu_0 H_z\,=\,0$, $T_c$). For finite $\mu_0 H_z$, the critical endpoint becomes strain controlled, enforcing a symmetric entropy landscape around the maximum at $\varepsilon_{xy}\,=\,0$ (see the inset in Fig.~\ref{fig:3}). Consequently, the ECE contribution is negative for compressive strain and grows with increasing $\mu_0 H_z$, as the field effectively renders $\lambda$ field dependent. Importantly, the ECE magnitude related to the AM term $\lambda$ can reach experimentally resolvable magnitudes even for moderate $\lambda$, since the Grüneisen parameter $\Gamma_{xy}\,=\,\eta_{xy}/T$ is expected to nearly diverge \cite{Bartosch10} at a finite-$T$ critical endpoint due to the entropy accumulation.

\begin{figure}
\includegraphics[width=\columnwidth]{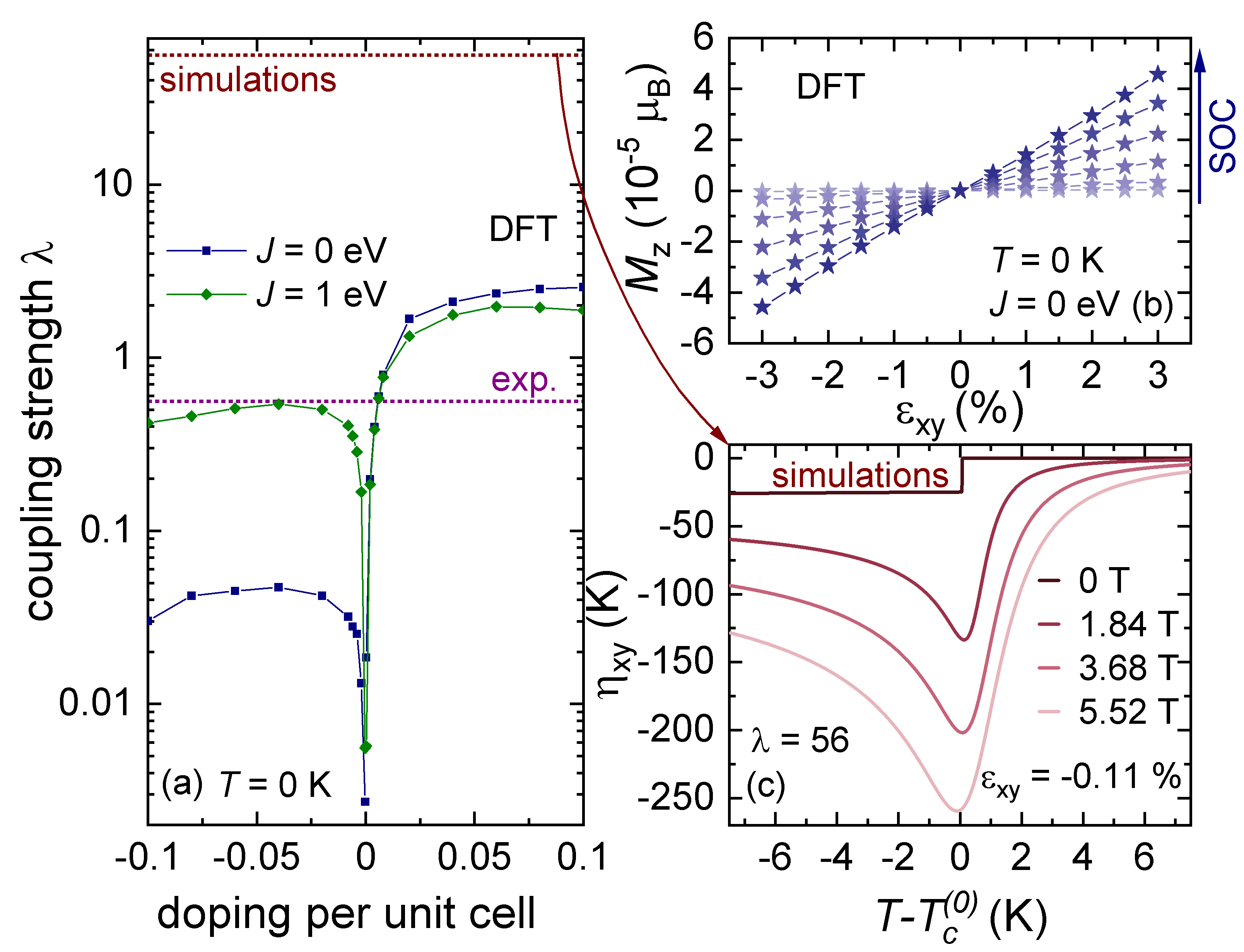}
\caption{\label{fig:4} (a) Magnitude of $\lambda$ per unit cell as a function of additional carrier concentration from DFT calculations. Introduction of $\sim 0.001$ holes per unit cell ($\sim 10^{19}\, {\rm cm}^{-3}$) in the DFT calculations provides agreement with the experimental observation (dotted pink line) for reasonable values of the Hund's coupling $J$. (b) Net magnetic moment $M_z$ per unit cell of MnF$_2$ as a function of $\epsilon_{xy}$ shear strain for different magnitudes of SOC calculated from DFT. Even though the spin group analysis allows the $M_z$ PZM response to be finite, our calculations indicate that the corresponding component $\lambda$ is suppressed at zero temperature because of the large band gap when SOC is not taken into account. (c) Simulations of $\eta_{xy}$ vs $T-T_{c}^{\left(0\right)}$ for a higher value of $\lambda$ [dotted red line in (a)], which may be realized in other materials with metallic character or larger SOC [see (b) and End Matter]. The predicted $\eta_{xy}$ values should be readily observable in experiments over a wider temperature range, enabling the direct measurement of the AM susceptibility.}
\end{figure}

Next, to gain insight into the microscopic origin and magnitude of $\lambda$, we link it to the piezomagnetic (PZM) response  obtained from first-principles DFT calculations [see Figs.~\ref{fig:4}\,(a) and \ref{fig:4}\,(b)]. Spin-group analysis allows only one PZM response to be nonzero without spin-orbit coupling, given by $M_z = -\partial f / \partial B_z = \lambda \mu_B \Phi \varepsilon_{xy}$ (see End Matter, Sec.\,\ref{sec:PZM}). Using that by convention $\Phi \approx 1$ deep in the ordered state, we estimate $\lambda$ from the calculated PZM response.

In the DFT calculations (see Supplemental Material, Sec.\,\ref{sec:DFT-methods}), the AM phase of MnF$_2$ with magnetic moments parallel to the crystallographic $c$ axis is considered, and self-consistent calculations are repeated for varying magnitudes of $\epsilon_{xy}$ strain to obtain the magnitude of the induced unit cell magnetic dipole moment along the $c$ axis. Interestingly, while spin-group analysis predicts that PZM induced by $\varepsilon_{xy}$ [see Fig.~\ref{fig:4}\,(b)] should be allowed in the AM phase without spin–orbit coupling (SOC)~\cite{Etxebarria2025}, DFT calculations show that it is strongly suppressed in the absence of SOC at zero temperature. This can be explained by the fact that MnF$_2$ is an insulator, and it is not possible to induce a magnetic moment in fully occupied bands without exciting electrons across the band gap \cite{Khodas25, Buiarelli2025}. This is supported by the observation that increasing electronic temperature enhances the PZM coefficient in DFT [see End Matter, Fig.~\ref{fig:theory}(a)].

 However, our DFT calculations for pristine MnF$_2$ with finite SOC consistently underestimate the experimentally observed $\lambda$, and introducing reasonable values of the Hubbard $U$ or Hund’s coupling $J$ does not alter this result [see Fig.~\ref{fig:theory}(b)]. A potential explanation for the discrepancy between calculations and experiment could be finite-temperature effects. It was recently proposed that thermally excited magnons also contribute to $\lambda$ as $T_c$ is approached from below in insulating antiferromagnets~\cite{Yershov24,Khodas25}.  However, in this case a strong temperature dependence of $\lambda$ is expected, which was not observed in an earlier measurement of the PZM in MnF$_2$ \cite{Bar19}.

In the following we show that another way to reconcile this disagreement is to take into account the presence of additional electrons or holes in the system. Although our samples are of high quality, we cannot fully exclude the presence of defects or slight off-stoichiometry. For example, even a one part in $10^3$ off-stoichiometry leads to $\sim 10^{19}\,{\rm cm}^{-3}$ electrons or holes, which is much higher than obtained through thermal excitations across the gap (note that, for piezomagnetism, what matters is whether the additional charges are spin polarizable, not whether they are itinerant; in insulating MnF$_2$, off-stoichiometry likely results in localized charges). In order to simulate this effect, we repeated our DFT calculations with added electrons or holes and extracted $\lambda$ [see Fig.~\ref{fig:4}(a)]. Both electrons and holes initially enhance $\lambda$ by orders of magnitude before it shows a saturationlike behavior.  Even though the sign of the coefficient is the same for either type of doping, there is strong electron-hole asymmetry in the size of response, which is particularly pronounced at $J\,=\,0$. This can be understood by considering which sites the carriers occupy in the DFT calculations [see Fig.~\ref{fig:theory}(c)]. Doped electrons occupy the widest $t_{2g}$ bands formed by $d$~orbitals elongated along the Mn chains along the $c$ axis. The spin density is nonzero on both opposite spin Mn ions, and, hence, their net magnetic moment is canceled to a large extent. Doped holes, on the other hand, occupy the narrower $e_g$ bands on top of the valence band, and the DFT results indicate that most of them occupy only one of the Mn ions' orbitals under strain, hence giving rise to an enhanced net magnetic moment. Introducing additional on-site Hund's coupling $J$ reduces this electron-hole asymmetry but does not completely eliminate it. In either case, we estimate that about 0.001 holes per Mn is sufficient to obtain results that agree with the experimental observation. Note, however, that the magnitude of this effect is meaningful only as an order of magnitude because of the tendencies of DFT to overestimate magnetic order in partially filled orbitals and to underestimate electron localization. 

These insights highlight that even very small deviations from stoichiometry can significantly affect the magnitude of $\lambda$ in AM insulators. For MnF$_2$, where only a single prior measurement exists for this combination of strain and field~\cite{Bar88} (see End Matter, Sec.\,\ref{sec:PZM}), this sensitivity likely rationalizes remaining discrepancies of earlier studies with our results, in addition to experimental factors such as background signals.

 
\textit{Summary and outlook - } In this Letter, we establish a direct thermodynamic measurement of the octupolar symmetry of the AM order parameter of MnF$_2$. By exploiting the elastocaloric effect as an exceptionally sensitive probe of the strain and field dependence of the AM free energy, we determine the cusp-shaped dependence of the crossover scale $T^*$ at the AM phase transition as a function of the field conjugate to the AM order parameter. In this way, the underlying PZM makes it possible to observe the otherwise hidden AM order parameter experimentally, even in systems, such as MnF$_2$, where the PZM coupling is relatively weak. 

Taken together, the present results open the way to applying this novel ECE methodology to other AM candidates with varying $\lambda$.  In AMs with a 2 orders of magnitude larger $\lambda$, substantial ECE signatures at finite $\hat{h}$ -- of the type shown in Fig.~\ref{fig:4}(c) -- should emerge. In this case, therefore, ECE measurements should not only enable measurements of the crossover lines and of the entropy accumulation close to $T_c$, but should also allow the determination of the AM susceptibility for $T\,\gg\,T_c$ \cite{Chakarborty2025,Steward25,Ye24}. Since we suggest from our DFT results that $\lambda$ can be orders of magnitude larger in AM metals, where a band gap is absent, ECE measurements should be particularly powerful in metallic $d$-wave AMs.


Importantly, the finite SOC in real materials will also extend the reach of ECE measurements beyond $d$-wave octupolar AMs. This is because in higher-order $g$-wave AMs such as hematite \cite{Verbeek24} and MnTe and CrSb, even though spin-group symmetries forbid intrinsic octupole moments, SOC necessarily induce them \cite{Fernandes2024_AM,Jaeschke2025,Buiarelli2025}, thereby leading to PZM responses \cite{Andratskil67,Aoy24}.

Beyond uncovering the thermodynamic signatures of unconventional symmetry breaking of multipolar order in AMs, the prediction of resolvable elastocaloric effects offers a compelling route to access AM fluctuations near putative quantum critical points \cite{Steward25,Chakarborty2025}, akin to studies of nematic quantum criticality in the pnictides \cite{Ikeda21}. This positions the ECE approach as a powerful probe of e.g. superconductivity \cite{Mazin25} induced by AM (i.e., multipolar) quantum critical fluctuations \cite{Wu2025}.

\textit{Acknowledgments - }We thank Bernd Wolf and Wolf Aßmus for kindly providing the MnF$_2$ crystal and Jan Priessnitz, Libor Smejkal and Charles R. W. Steward for useful discussions. Financial support from the Max Planck Society (R. O., H. M. L. N., A. P. M., and E. G.)  is gratefully acknowledged. In addition, R. O., H. M. L. N., A. P. M., J. S., and E. G. gratefully acknowledge the funding through the Deutsche Forschungsgemeinschaft (DFG, German Research Foundation) through Grant No. TRR 288—422213477 Elasto - Q - Mat (Projects No. A10, No. A13 and, No. B01). Research in Dresden benefits from the environment provided by the DFG Cluster of Excellence ctd.qmat (EXC 2147, Project ID No. 390858490). R. M. F. was supported by the Air Force Office of Scientific Research under Award No. FA9550-21-1-0423. R. M. F.
also acknowledges a Mercator Fellowship from the German Research Foundation (DFG) through Grant No. TRR 288,
422213477 Elasto-Q-Mat.
L. B. and T. B. were supported by the National Science Foundation CAREER Grant No. DMR-2046020 and also partially supported by the National Science Foundation through the University of Minnesota MRSEC under Award No. DMR-2011401. 

\textit{Data availability - }
The data that support the findings of this article are openly available~\cite{Data26}.

\newpage
\clearpage

\section*{End Matter}

\subsection{Elastocaloric data at different fields and strains}
\label{sec:parameters}

\begin{figure}[h!]
\includegraphics[width=0.48\textwidth]{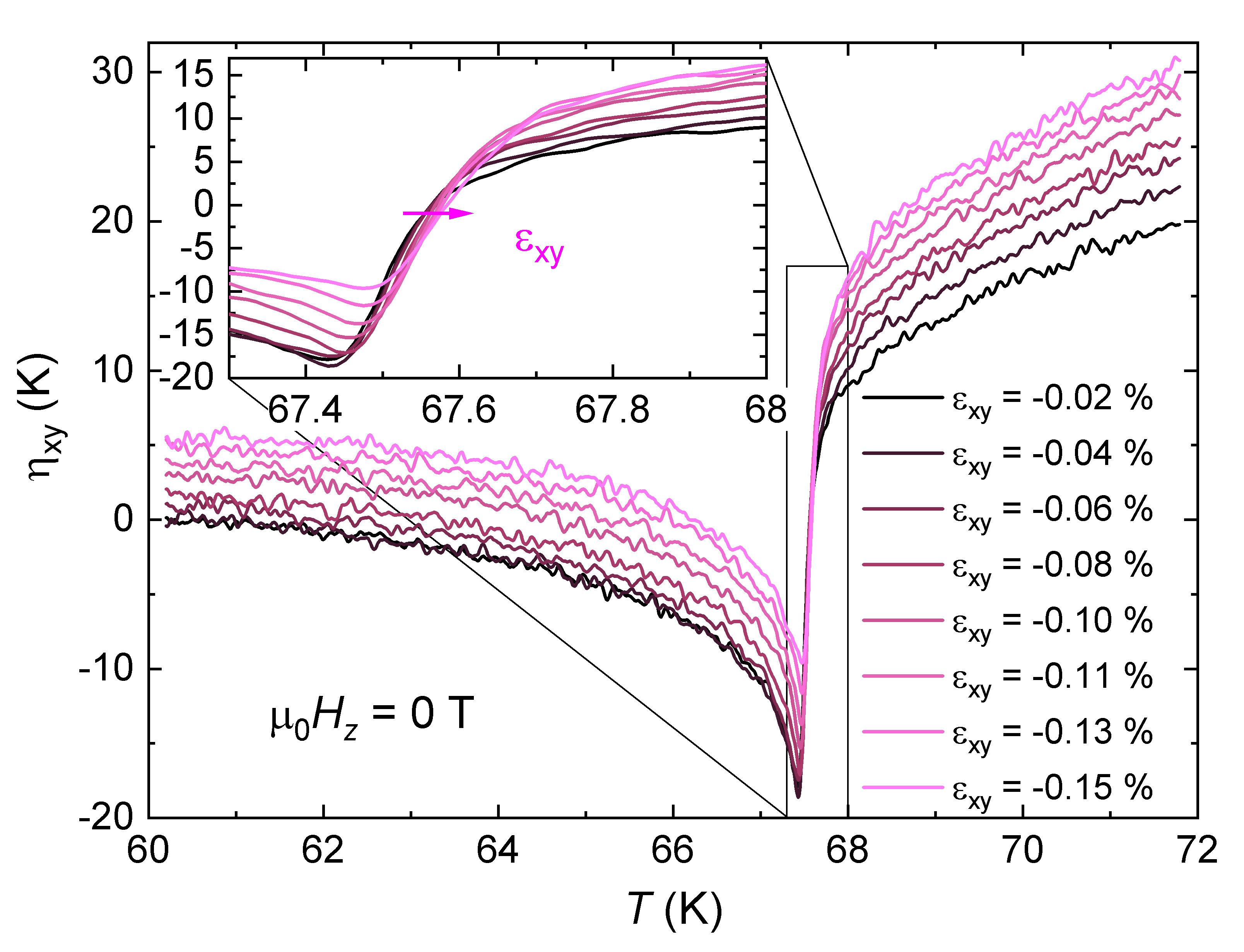}
\caption{\label{fig:eps_data}Elastocaloric effect, $\eta_{xy}=\Delta T/\Delta \varepsilon_{xy}$, of MnF$_2$ vs temperature $T$, at different constant strains and zero magnetic field. The inset shows an enlarged view of the data around the phase transition. The pink arrow indicates that the phase transition shifts to higher temperatures with increasing compression (negative strains). The change in the magnitude of the elastocaloric effect at high temperatures likely results from small strain-induced changes of the phononic contributions and is denoted by $\eta_n$ in the main text}.
\end{figure}

In the following, we present all experimental data of the ECE in MnF$_2$ as a function of $T$ at different $\varepsilon_{xy}$ and $\mu_0 H_z$, that we used to infer the $T^*-T_c^{(0)}$ data in Fig.\,\ref{fig:1} of the main text. Figure \ref{fig:eps_data} shows $\eta_{xy}$ of MnF$_2$ vs $T$ at zero magnetic field and at different compressive strains $\varepsilon_{xy}$ up to $-0.15\%$. The data reveal a clear steplike feature around $T\,\approx\,$67.5\,K, characteristic for a second-order phase transition, as expected for MnF$_2$ at zero strain. To better illustrate the strain-induced shift of the transition temperature, the phase transition anomaly is shown on enlarged scales in the inset in Fig.~\ref{fig:eps_data}. We find that the transition temperature shifts linearly to higher temperatures with increasing compression (see also Fig.~\ref{fig:TvsEps} in Supplemental Material) and follows $T^\prime_c\,=\,T_c^{(0)} + a_1 \varepsilon_{xy}$ with a slope of $a_1\,=\,$-59(3)\,K. The ECE jump remains essentially unchanged at different $\varepsilon_{xy}$, as expected for a constant $a_1$ and a constant specific heat $C_\varepsilon$, following the Ehrenfest relation

\begin{equation}
\frac{\textnormal{d}T_c^{(0)}}{\textnormal{d}\varepsilon_{xy}} = \frac{\Delta \left[C_\varepsilon \eta_{xy}\right]}{\Delta C_\varepsilon}.
\label{eq:Ehrenfest}
\end{equation}

\begin{figure*}
  \centering
    \includegraphics[width=1\textwidth]{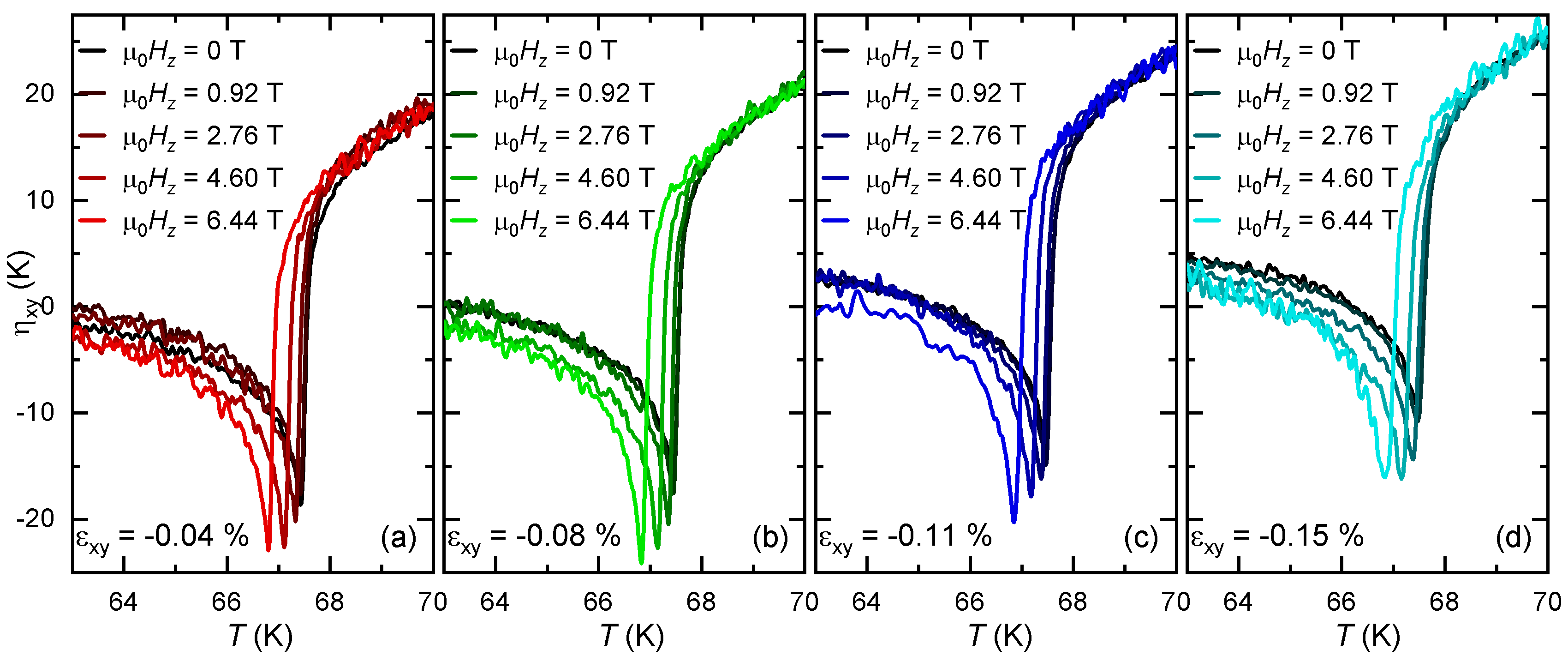}
    \caption{\label{fig:H_data}Elastocaloric effect, $\eta_{xy}=\Delta T/\Delta \varepsilon_{xy}$, of MnF$_2$ vs temperature $T$, taken at constant strains [(a) $\varepsilon_{xy}=-0.04\,\mathrm{\%}$, (b) $\varepsilon_{xy}=-0.08\,\mathrm{\%}$, (c) $\varepsilon_{xy}=-0.11\,\mathrm{\%}$, and (d) $\varepsilon_{xy}=-0.15\,\mathrm{\%}$] at different magnetic fields (0 -- 6.44 T).}

  \vspace{2mm}

    \includegraphics[width=1\columnwidth]{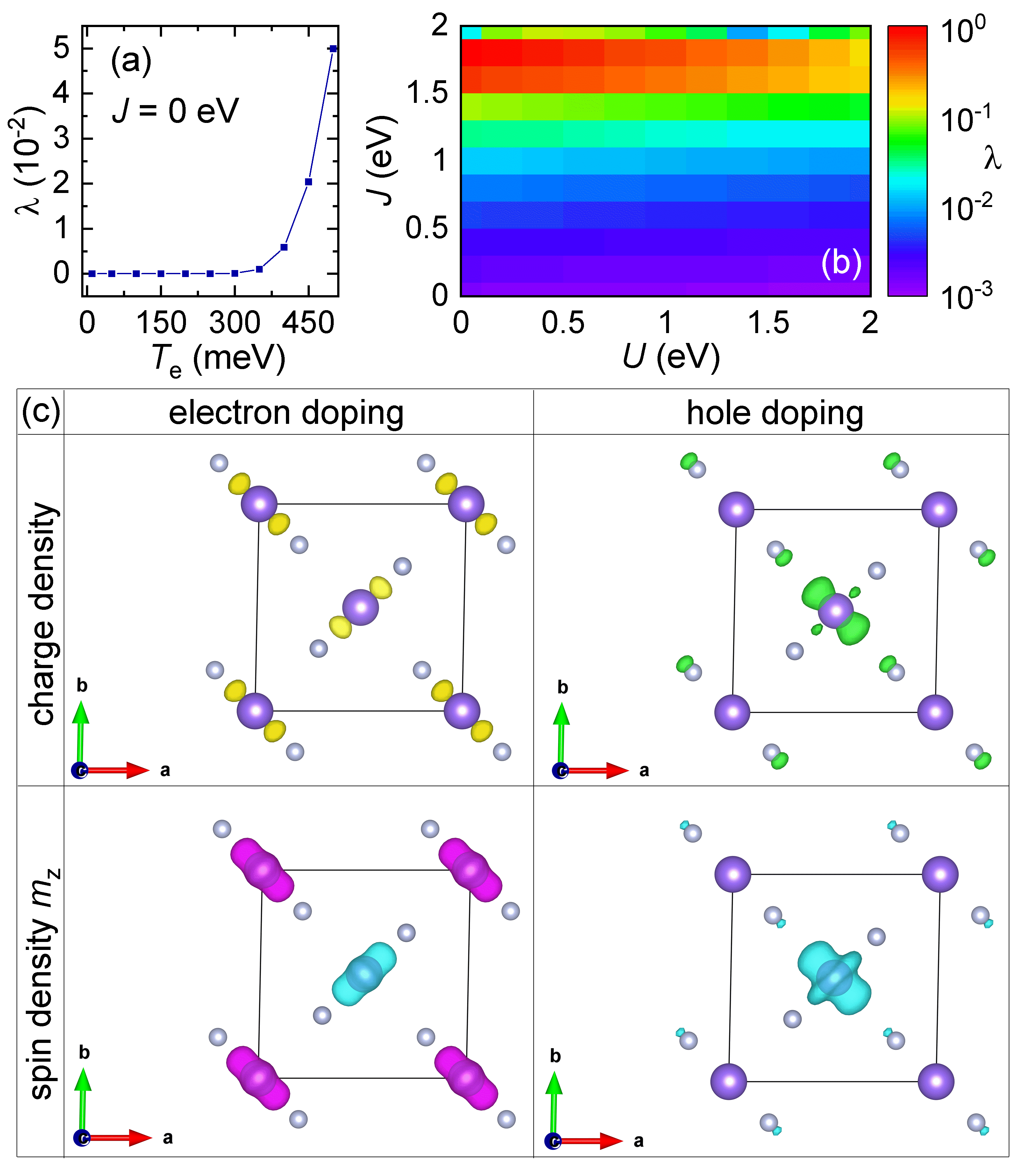}
    \caption{\label{fig:theory} (a)  Calculated  $\lambda$ as a function of electronic temperature, extracted from calculations of the piezomagnetism through DFT calculations of fully stoichiometric MnF$_2$. For large enough electronic temperatures that give rise to considerable number of electron-hole pairs to be excited, $\lambda$ is strongly enhanced. (b) When an on-site + $U$ correction is introduced in the DFT calculations, $\lambda$ shows a minor dependence on the value of $U$ used but a stronger dependence on the Hund's coupling $J$. Nevertheless, physically meaningful values of $J$ do not lead to large enough $\lambda$ values that agree with the experimental observations. (c) Isosurfaces of the doping-induced changes in the charge densities (green, positive change; yellow, negative change) and spin densities (pink, positive change; cyan, negative change) under electron and hole doping, shown around the Mn ions (purple) and F ions (gray).} 
\end{figure*}

Next, we present the ECE at $\mu_0 H_z$ for four selected strain values between $-0.04\,\%$ and $-0.15\,\%$, as shown in Figs.~\ref{fig:H_data}\,(a)–\ref{fig:H_data}\,(d). For each strain, the ECE feature shifts to lower temperatures with increasing $\mu_0 H_z$ up to 6.44\,T, and this suppression is essentially symmetric in field (see Supplemental Material). At zero strain, the field dependence of the critical temperature is well described by $T^\prime_c = T_c^{(0)} + a_2 (\mu_0 H_z)^2$ with $a_2 = -0.01585(14)\,$K/T$^2$ (see Supplemental Material, Sec.\,\ref{sec:globalfit}), in excellent agreement with earlier reports of the $T$–$\mu_0 H_z$ phase diagram \cite{Shapira69}.

\subsection{Connection to piezomagnetic coefficients}
\label{sec:PZM}




In Ref.\,\cite{Bar88}, the PZM coefficient of MnF$_2$ was reported as the magnetization induced by stress, $\sigma_{jk}$: 

\begin{equation}
    M_i=\Lambda_{ijk}\sigma_{jk}.
    \label{eq:PZM2}
\end{equation}

The stress $\sigma_{110}$ can be connected to the strain $\varepsilon_{xy}$ using the appropriate Young's modulus $C_0$ and the Poisson ratios via $\sigma_{110}\,=\, \frac{2 C_0}{1+\nu}\varepsilon_{xy}$. Together with the definition of the PZM in the main text, this allows us to express $\lambda$ in terms of the measured $\Lambda$:

\begin{equation}
\lambda\,\approx\,\frac{\Lambda}{\mu_B \Phi} \frac{2 C_0}{1+\nu}.
    \label{eq:comparison}
\end{equation}

In Table~\ref{tab:piez}, we list the previously reported value of $\Lambda_{zyx}$, which is the PZM coefficient allowed by symmetry only, as well as the one of $\Lambda_{yxz}$, which is allowed only in the presence of SOC, as well as the corresponding $\lambda_{ijk}$ values using the assumption that $\Phi\,\approx\,1$. This assumption is likely violated, especially near $T_c^{(0)}$ where $\Phi$ becomes small, leading to an underestimation of $\lambda$ according to Eq.~\eqref{eq:comparison}.

\begin{table}
\begin{ruledtabular}
\begin{tabular}{cdcr}
\textrm{$\Lambda_{ijk}$}&
\multicolumn{1}{c}{\textrm{$\lambda_{ijk}$}}&
\textrm{$T$}&
\textrm{Ref}\\
\colrule
 $\Lambda_{zyx}=5.82e^{-7}\,\mathrm{\mu_B/MPa}$ & 0.176 & 60 K&\cite{Bar88}\\
  $\Lambda_{yzx}=8.15e^{-7}\,\mathrm{\mu_B/MPa}$ & 0.246 & 20 K&\cite{Bor60}\\
 $\Lambda_{ave}=1.20e^{-7}\,\mathrm{\mu_B/MPa}$ & 0.036 & 50 K&\cite{Kom25}\\
\end{tabular}
\end{ruledtabular}
\caption{\label{tab:piez}%
Literature values of the piezomagnetic coefficient per unit cell of MnF$_2$ and the corresponding AM coupling constant $\lambda$ calculated by means of Eq.\, \ref{eq:comparison}. Additionally, the temperature at which the piezomagnetic coefficient was measured is stated. Only $\Lambda_{zyx}$ is allowed by symmetry, whereas $\Lambda_{yzx}$ is allowed only in the presence of SOC. $\Lambda_{ave}$ refers to a measurement on a polycrystal.}
\end{table}

The DFT-based calculation results for $\lambda$ of pristine MnF$_2$ are shown in Figs.~\ref{fig:theory} (a) and \ref{fig:theory} (b) as a function of the electronic temperature $T_e$ for $J\,=\,0\,$eV (a) and as a function of $J$ and $U$ at $T\,=\,0\,$K (b). In addition, Fig.~\ref{fig:theory}\,(c) shows the change in the charge and spin densities upon electron and hole doping, respectively (see the main text, Fig.\,\ref{fig:4}).

\newpage
\clearpage

\setcounter{figure}{0}
\renewcommand{\thefigure}{S\arabic{figure}}
\setcounter{equation}{0}
\renewcommand{\theequation}{S\arabic{equation}}

\pagestyle{empty}


\subsection{Further background information on the theoretical analyses}

\subsubsection{The crossover temperature in mean-field and other universality classes}
\label{sec:crossover-theory}

In analogy to the case of an Ising ferromagnet in finite field, in which ferroically ordered dipoles couple bilinearly to a magnetic field, we can treat the  ordered octupoles in an AM as ferroically interacting pseudospins with conjugate field $\hat{h}=\mu_B\varepsilon_{xy} \mu_0H_z$. To this end, we consider the following Hamiltonian

\begin{equation}
    \mathcal{H}=-J\sum_{<ij>}\tau_i^z\tau_j^z -\lambda \hat{h} \sum_i\tau_i^z,
    \label{eq:Hamiltonian}
\end{equation}
where $\tau^z_i = \pm 1$ is the Ising variable characterizing the ferrooctupolar degree of freedom and $J>0$ describes the interaction between the octupolar degrees of freedom. In order to get a sense of the conjugate field-induced AM order, we start with a simple mean-field analysis of Eq.~\eqref{eq:Hamiltonian}, while $\lambda$ is a dimensionless measure of the piezomagnetism that is responsible for the combined strain and field effects exploited here.
We expand the equation of state, derived from the Hamiltonian in Eq.~\eqref{eq:Hamiltonian}, in the vicinity of $T_c^{(0)}$ with $k_BT_c^{(0)}=zJ$ the transition temperature in the absence of strain or magnetic field and $z$ the number of nearest neighbors. 
\begin{align}
    \Phi=\textnormal{tanh}\left(\frac{k_BT_c^{(0)}\Phi+\lambda \hat{h}}{k_BT}\right)\nonumber \\
     \approx \frac{1}{T}\left(T_c^{(0)}\Phi+\frac{\lambda}{k_B}\hat{h}-\frac{T_c^{(0)}}{3}\Phi^3\right)
\end{align}
where $\Phi=\langle\tau^z\rangle$. This leads to the following cubic equation 
\begin{equation}
    (T-T_c^{(0)})\Phi+\frac{T_c^{(0)}}{3}\Phi^3=\frac{\lambda}{k_B}\hat{h}.
    \label{eq:eos_mf}
\end{equation}
with solution
\begin{equation}
\Phi=\left(\frac{3\lambda \hat{h}}{k_BT_c^{(0)}}\right)^{1/3}\psi\left(\frac{T-T_c^{(0)}}{\left(\frac{\lambda}{k_B}\hat{h}\right)^{2/3}\left(T_c^{(0)}\right)^{1/3}}\right)
\end{equation}
where  $\psi(x)$ is the solution of  $3^{1/3}x \psi+\psi^3=1$.
The crossover induced by the conjugate field  occurs whenever the argument of the function $\psi(x)$ is of order unity, i.e. for 
\begin{equation}
    T^*= T_c^{(0)}+T_c^{(0)}\left(\frac{\lambda \lvert \hat{h} \rvert}{k_B T_c^{(0)}}\right)^{2/3},
    \label{eq:Tstarinmeanfield}
\end{equation}
where the absolute value of the conjugate field was introduced to account for negative values of $H_z$ or $\varepsilon_{xy}$.
It is convenient to rewrite the condition for $T^*$ in dimensionless form, where we inserted our explicit expression for $\hat{h}$:
\begin{equation}
    \frac{T^*-T_c^{(0)}}{T_c^{(0)}}=\left(\frac{\lvert\mu_0H_z\varepsilon_{xy}\rvert}{B^*}\right)^{2/3}.
\end{equation}
where we introduced the magnetic field scale 
\begin{equation}
  \mu_BB^*=k_BT_c^{(0)}/\lambda, 
\end{equation}
i.e., the magnetic field that yields a piezomagnetic energy of order the transition temperature. 
If we use our above values for $\lambda$ and $T_c^{(0)}$ it follows $B^*= 179\,\mathrm{T}$, illustrating that the piezomagnetic coupling, while allowing for symmetry-sensitive effects, is comparatively small.

This mean-field analysis allows for a very natural generalization to non-mean-field behavior that includes critical fluctuations. This is easiest done using the dimensionless temperature scale
\begin{equation}
    t\equiv\frac{T-T_c^{(0)}}{T_c^{(0)}}
\end{equation}
and a dimensionless conjugate field
\begin{equation}
    b\equiv\frac{\lvert\mu_0H_z\varepsilon_{xy}\rvert}{B^*}=\frac{\lambda\mu_B\lvert\mu_0H_z\varepsilon_{xy}\rvert}{k_BT_c^{(0)}}.
\end{equation}
Then the generic scaling relation of an Ising order parameter holds,
\begin{equation}
    \Phi(t, b) =b^{1/\delta}\Psi\left(tb^{-\frac{1}{\beta\delta}}\right)
\end{equation}
with  universal scaling function $\Psi(x)$. The crossover between the $b$-dominated and $t$-dominated behavior occurs when the argument of the scaling function is of order unity, i.e., $T^*=b^{1/(\beta\delta)}$. Hence this corresponds to following expression for the crossover temperature that generalizes Eq.~\eqref{eq:Tstarinmeanfield}
\begin{equation}
    \frac{T^*-T_c^{(0)}}{T_c^{(0)}}=\left(\frac{\lvert\mu_0H_z\varepsilon_{xy}\rvert}{B^*}\right)^{1/(\beta \delta)}.
\end{equation}
In mean-field theory $\beta =\frac{1}{2}$ and $\delta = 3$, which recovers the above result
with $1/(\beta\delta)_\textrm{MF} = 2/3$. For the three-dimensional (3D) Ising model $\beta\,\approx\,0.31$ and $\delta\,\approx\,5$, which gives instead a value of $1/(\beta\delta)_\textrm{3D}\,\approx 0.64$. For the two-dimensional (2D) Ising
model $\beta=1/8$ and $\delta=15$. Then the exponent is $1/(\beta\delta)_\textrm{2D}=8/15\approx 0.533$.

\subsubsection{Strain and field coupling beyond piezomagnetism}
\label{sec:strain-decomposition}

The equation of state given in Eq.~\eqref{eq:eos_mf} follows from the Landau free-energy expansion~\cite{Steward23,Chakarborty2025,Steward25}
\begin{equation}
     f=\frac{k_B}{2}\left(T-T_c^{(0)}\right)\Phi^2+ \frac{u}{4}\Phi^4-\lambda \hat{h}\Phi.
     \label{eq:free-energy}
\end{equation}
with $u=k_BT_c^{(0)}/3$ and $f=F/N$ has been normalized by he number of unit cells $N$.

In order to induce a $\varepsilon_{xy}$ strain, we apply a $\sigma_{110}$ stress. The induced $\varepsilon_{110}$ can be decomposed into its irreducible components \cite{Ike18}, i.e., a pure shear strain $\varepsilon_{xy}$ and a fully symmetric strain $\frac{1}{2}(\varepsilon_{xx}+\varepsilon_{yy})$, as shown in Fig. \ref{fig:2}\,(a) of the main text. Specifically, the strains are given by
\begin{equation} \label{eq:decomposition}
\begin{aligned}
    \frac{1}{2}(\varepsilon_{xx}+\varepsilon_{yy}) &= \frac{1}{2} (1 - \nu) \varepsilon_{110} \\
    \varepsilon_{xy} &= \frac{1}{2} (1 + \nu) \varepsilon_{110},
\end{aligned}
\end{equation}
with $\nu\,=\,-\varepsilon_{1\,\bar{1}\,0}/\varepsilon_{1\,1\,0}$ the in-plane Poisson ratio. 

Based on symmetry considerations, additional coupling to symmetric strain and higher-order coupling terms to the field are allowed in the free energy. This leads to symmetric strain and quadratic magnetic field dependent changes of $T_c^{(0)}$ in the first term of Eq.~\eqref{eq:free-energy}: 
\begin{equation}
    T_c^{(0)}\rightarrow T_c=T_c^{(0)}+a_0 \frac{1}{2}(\varepsilon_{xx}+\varepsilon_{yy})+a_2 (\mu_0H_z)^2
\end{equation}
The effect of the additional coupling terms is that the symmetry-preserving $\frac{1}{2}(\varepsilon_{xx}+\varepsilon_{yy})$ strain can change the ordering temperature of an AM linearly, while the transition temperature can shift quadratically in field even at zero strain. For further comparison with experiments it is useful to rewrite this free-energy model in terms of the shear strain $\varepsilon_{xy}$ with the help of Eq.~\eqref{eq:decomposition}. The result is given in Eq.~\eqref{eq:free-energy-expanded} of the main text. Here the proportionality factors between $\varepsilon_{110}$ and the symmetric as well as the shear strain, which depend on $\nu$ and are of the order one, have been absorbed into $a_1$ and $\lambda$, respectively.

\subsubsection{Further details on the simulations of the ECE}
\label{sec:simulations}

\begin{figure}
\includegraphics[width=0.48\textwidth]{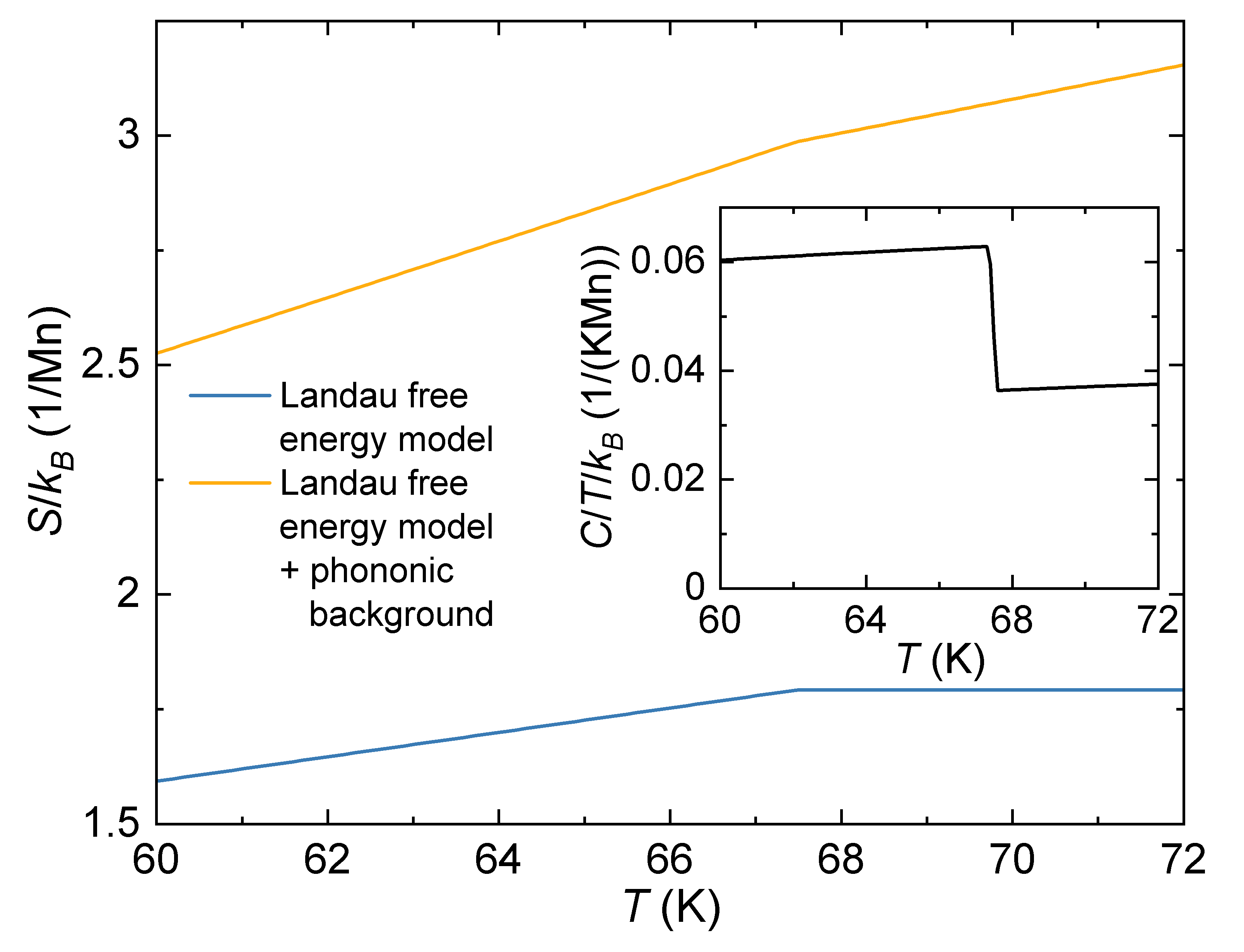}
\caption{\label{fig:modelCT} Entropy, $S/k_B$, vs. temperature, $T$, at zero strain and zero field obtained in our simulations from the free energy Eq.~\eqref{eq:free-energy-expanded} (blue line) and with an additional phononic contribution that matches the one of MnF$_2$ (orange line). The inset shows the heat capacity, $C/T$ in units of $k_B$, at zero strain for the model including phononic contributions. The entropy axis is shifted and does not show the true zero.}
\end{figure}

In Fig.~\ref{fig:modelCT}, we show the entropy, $S$, and the heat capacity, $C/T$, at zero strain obtained from our model simulations. The entropy was calculated using Eq.~\eqref{eq:free-energy-expanded} of the main text with prefactors that were chosen such that the high-temperature entropy per spin is ln(6) for the $S\,=\,5/2$ of Mn$^{2+}$ (see blue line). Since we experimentally determine the ECE with respect to the total heat capacity of the system, we next added a phononic entropy, that was based on previous estimates of the phonon contribution in literature \cite{Boo76}. The total entropy is shown by the orange line. The heat capacity, $C/T$, for the model including phononic contributions is shown in the inset. The ratio $\Delta C/C_n$, with $\Delta C$ being the jump size and $C_n$ the value of the heat capacity just above the transition, is in good agreement with the one obtained in experimental heat capacity data on MnF$_2$ \cite{Ike78}. We note that the phononic entropy was kept constant throughout our simulations, i.e., we did not impose a strain dependence.

\subsubsection{Simulation results for tensile strains}

\begin{figure}
\includegraphics[width=0.48\textwidth]{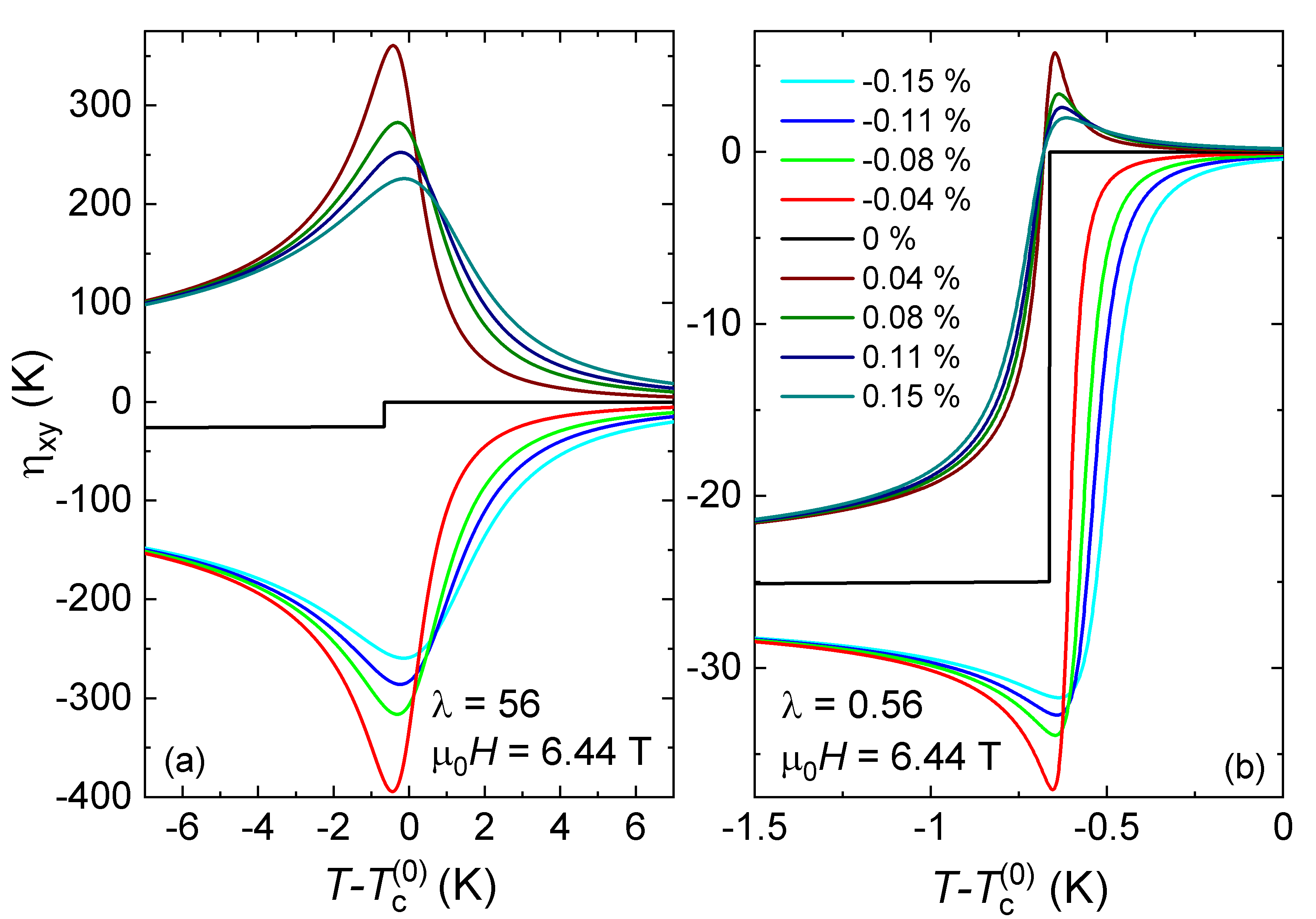}
\caption{\label{fig:sim_strain} Simulations of the elastocaloric effect, $\eta_{xy}$, as a function of relative temperature, $T-T_c^{(0)}$ by means of a Landau free energy expansion with respect to the octupolar order parameter (Eq.~\eqref{eq:free-energy-expanded} of the main text) at a constant field of $\mu_0 H_z\,=\,6.44\,\mathrm{T}$ for (a) a strong octupolar coupling constant $\lambda\,=\,56$ and (b) a weak coupling constant $\lambda\,=\,0.56$. Note the different scales of the temperature and ECE axes in (a) and (b).}
\end{figure}

In the main text (cf. Fig.\,\ref{fig:3}), we argued that the ECE magnitude in finite field and strain close to $T_c^{(0)}$ is governed by the near ``divergence" of the Grüneisen parameter, $\Gamma_{xy}$, of the underlying ferro-octupolar critical point. Here, we complete the discussion by showing results of simulations under tensile strain. Unfortunately, we were unable to study MnF$_2$ experimentally under tension, since the crystals broke at very small tensile strains below 0.02\,\%. Thus, we focus on the results of simulations here.

The ferro-octupolar critical point, as defined by the parameter $\lambda$ in the free energy in Eqs.\,\eqref{eq:free-energy} and \eqref{eq:free-energy-expanded} (main text), is strain-controlled when the field is finite. As a consequence, we expect the features in our simulations, that arise from $\lambda$ to change sign under tensile strain (see discussion in the main text). The simulations shown in Fig.~\ref{fig:sim_strain} (a) for $\lambda\,=\,56$ and (b) $\lambda\,=\,0.56$ show that this is indeed the case. For $\lambda\,=\,56$, i.e., when AM thermodynamic signatures dominate, the sign change of $\eta_{xy}$ when going from tension to compression is clearly visible. Even for $\lambda\,=\,0.56$, where the effects are more subtle, AM signatures remain discernible. In this case, the dominant contribution to $\eta_{xy}$ stems from $a_1$, as reflected in the large jump of $\eta_{xy}$ at zero strain, but additional contributions near $T_c^{(0)}$ in finite fields are positive (negative) under tensile (compressive) strain.

\subsubsection{First-principles calculations}
\label{sec:DFT-methods}
We simulated MnF$_2$ using Density Functional Theory (DFT) with Projector Augmented-Wave (PAW) spinor wavefunctions, as implemented in \textsc{Abinit}~9.10.1~\cite{Torrent2008Apr, Marques2012Oct, Gonze2016Aug, Gonze2020Mar, Romero2020Mar}. The calculations employed JTH~v1.1 pseudopotentials~\cite{Jollet2014Apr} within the Local Density Approximation (LDA). A $\Gamma$-centered $16\times16\times24$ $k$-point grid was used, together with a plane-wave cutoff energy of 550~eV for the regular basis and 1100~eV for the spherical grid inside the augmentation regions. The fully relaxed crystal structure was used, with all Hellmann--Feynman forces minimized below 1~meV/\AA. The Hubbard and Hund terms were treated within the mean-field LDA+$U$ scheme in the fully localized limit~\cite{Amadon2008}. For most calculations, we used the insulator occupation scheme of \textsc{Abinit}, while for simulations as a function of electronic temperature $T_e$ and doping, we employed a Gaussian occupation scheme with a smearing temperature of 10~meV and physical temperature $T_e$~\cite{Verstraete2001}. To simulate the effect of doping, a compensating background charge was included using the \textit{cellcharge} flag of \textsc{Abinit}, such that the total charge density is given by $n_x(\mathbf{r}) = n(\mathbf{r}) + e x$, where $-e$ is the elementary charge and $x$ denotes the fraction of holes ($x>0$) or electrons ($x<0$) added per unit cell.

We apply the $xy$ strain $\epsilon$ to MnF$_2$ with the experimentally resolved structure, by setting the reduced (fractional) lattice vectors to be
\begin{equation}
    \begin{bmatrix}
    \sqrt{1+\epsilon^2} & \epsilon & 0 \\
    \epsilon & \sqrt{1+\epsilon^2} & 0 \\
    0 & 0 & 1 \\
    \end{bmatrix},
\end{equation}
where the diagonal terms make sure that the determinant of the matrix is 1, such that there is no volume change (no symmetric strain). After applying strain we let the internal atomic coordinates relax.

\subsection{Further details on the experiments}

\subsubsection{Methods}
\label{sec:exp-methods}

To study the response of MnF$_2$ to uniaxial stress, single crystals of MnF$_2$ were oriented along the tetragonal [1\,1\,0] direction. Subsequently, the samples were first polished by hand and then by Plasma Focused Ion Beam into a dumbbell shape (see e.g. Ref.\,\cite{Noa23}), in which the strain is concentrated in a neck of dimensions of approx.\,$700\,\mathrm{\mu m}\times140\,\mathrm{\mu m}\times140\,\mathrm{\mu m}$, with the long axis being the [1\,1\,0] direction. 

For the application of \textit{in situ} tunable uniaxial pressure, a piezoelectric-driven uniaxial pressure cell, similar to the one reported in Ref.\,\cite{Bar19}, was used. The applied stress, $\sigma_{110}$, and the resulting strain, $\varepsilon_{110}$, are measured through capacitive sensors installed in the cell. $\varepsilon_{110}$ was converted into $\varepsilon_{xy}$ using the reported elastic tensor, see below. A maximum $\varepsilon_{110}$ of around -0.25\,\% was reached in the total five sample studied. By measuring the stress-strain relationship \textit{in situ}, we verified that MnF$_2$ remained within the linear elastic regime for all the data reported in this work. For higher compression, signs of plastic deformation were observed in the $\sigma_{110}$ vs. $\varepsilon_{110}$ curve and features in the ECE became significantly broadened. As a result, only data taken in the elastic regime are considered in the present manuscript.

To determine the zero-stress point with maximum reliability, the sample was mounted on a two-part sample carrier (see, e.g., Refs.\,\cite{Jerzembeck22,Lieberich25}), as used in previous studies. This design ensures that the uniaxial pressure cell transmits force to the sample only when the two carrier pieces are in mechanical contact. As a result, the sample can be compressed but not placed under tension. The zero-stress point is readily identified from the applied force--displacement curve, which exhibits a distinct change in slope when the mechanical contact opens.

The strain values quoted in this work were calculated~\cite{Noa23} from the applied force, the sample's cross-sectional area, and the known Young's modulus of MnF$_2$ of $C_0\,=\,140\,$GPa \cite{Mel70}, that is essentially temperature-independent across experimental temperature range. To determine the applied force to the sample, the measured force was corrected for the small contribution of the sample carrier.

Low-temperature elastocaloric effect measurements \cite{Ike19,Li22,Ye24} in magnetic fields were performed by applying an a.c. strain to the sample through the piezoelectric actuators and measuring the resulting temperature change using a calibrated, insulated Chromel-AuFe thermocouple (Ref.\,\cite{Sto11}) attached to the top of the sample. In order to achieve best possible resolution, the two thermocouple leads were glued together to minimize the area between them and to avoid vibrations in field. The thermocouple voltage was amplified by both a low-temperature transformer as well as a low-noise high-temperature amplifier. The voltage signal was then read out using lock-in amplifiers. The frequency of the a.c. strain was chosen to be 39.13\,Hz to ensure quasi-adiabatic conditions (see Sec.\,\ref{sec:quasiadiabatic} below). A small a.c. strain, applied in addition to the d.c. strain and along the same direction, enables simultaneous tuning of the material via strain and probing of the elastocaloric response $\eta_{xy}(\varepsilon_{xy})$.

As a consequence of this experimental design, the ECE can only be measured at finite offset compression. At zero offset strain, the thermocouple readout is dominated by the opening of the mechanical contact in the two-part carrier: If the mechanical contact is open, the sample does not experience an a.c. strain, even if an a.c. voltage is applied to the piezoelectric actuators, and thus does not change its temperature and, as a result, no voltage should be induced in the thermocouple. The lock-in voltage readout, when the mechanical contact is opened, results from parasitic extrinsic signals, which we subtract from our data, taken when the mechanical contact is closed.

All measurements in a finite magnetic field were performed after field cooling. The temperature-dependent ECE data presented in the main text and in the End Matter correspond to data symmetrized with respect to positive and negative fields, i.e.,
$\frac{\Delta T}{\Delta \varepsilon}(H) = (\frac{\Delta T}{\Delta \varepsilon}(H) + \frac{\Delta T}{\Delta \varepsilon}(-H))/2$.
This procedure removes extrinsic contributions, for instance those arising from magnetic-flux changes within the thermocouple loop. The characteristic temperatures were extracted for positive and negative fields individually.

\subsubsection{Quasi-adiabaticity and absolute values of the elastocaloric effect}
\label{sec:quasiadiabatic}

\begin{figure}
\includegraphics[width=0.48\textwidth]{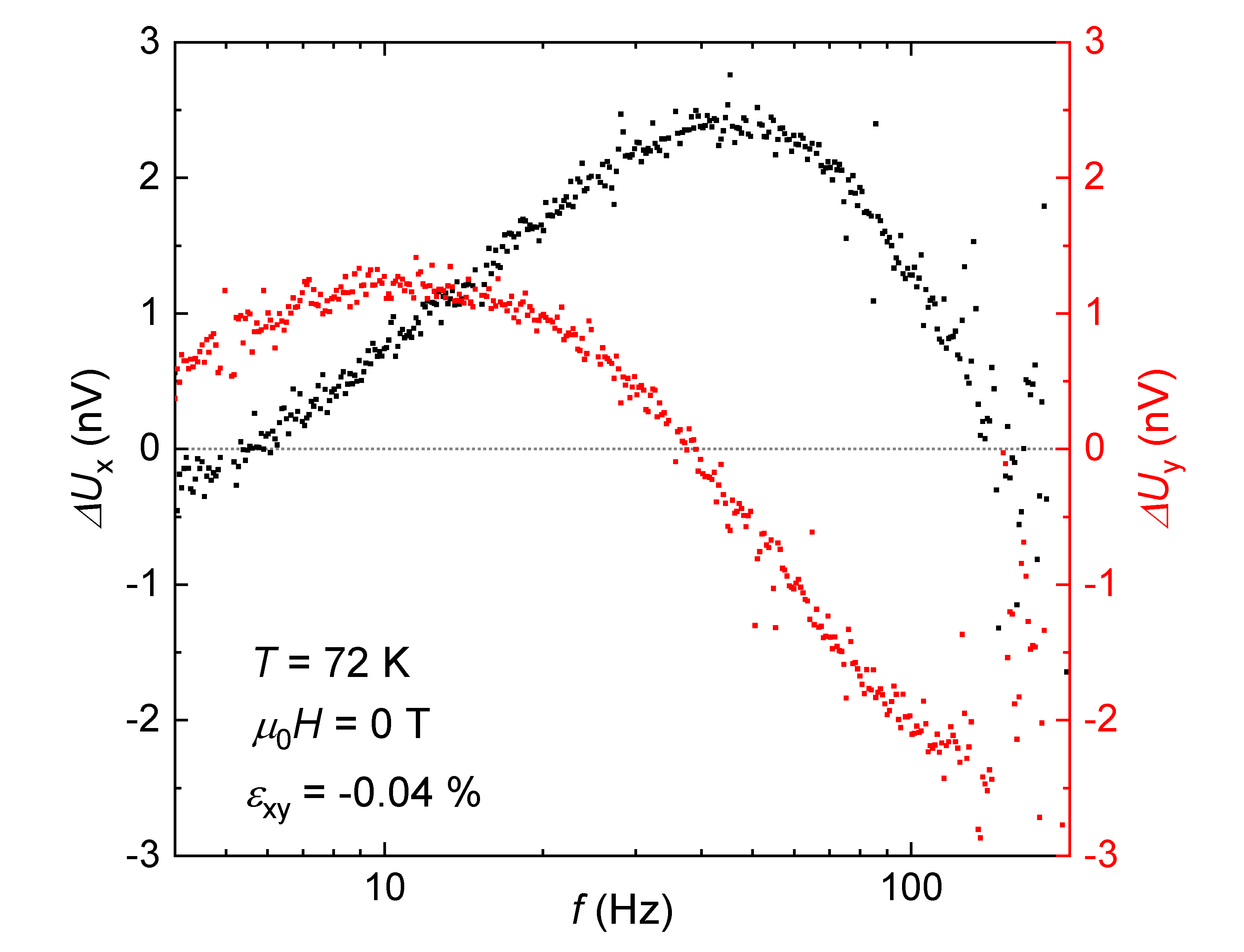}
\caption{\label{fig:XYvsF} The x- ($\Delta U_x$) and the y-component ($\Delta U_y$) of the voltage oscillation of the thermocouple, mounted on MnF$_2$, measured by the lock-in amplifier as a function of frequency. A background contribution measured in an open-gap configuration has been subtracted (see the method section for more details).}
\end{figure}

In order to measure the ECE under quasi-adiabatic conditions, the measurement frequency was chosen such that (i) it is high enough that as little heat as possible dissipates into the bath and (ii) it is small enough that the thermocouple is well thermalized. Experimentally, this frequency is determined from the maximum of the recorded thermocouple voltage as a function of frequency at fixed temperature and field (see Fig. \ref{fig:XYvsF}) \cite{Ike19}. In the present study, the optimal frequency was determined at 60\,K and 72\,K and selected fields up to 7.36\,T. For all data shown, i.e., data up to 6.44\,T, the optimal frequency did not change. Only for highest fields the optimal frequency changed. Thus, we use the 7.36\,T data only to determine the crossover temperature, but not in the discussion of absolute values of the ECE.

As noted in previous studies \cite{Straquadine20,Li22,Ike19,Lieberich25}, the absolute value of the ECE measured even at the optimal excitation frequency may be reduced compared to the intrinsic ECE, owing to heat flow into the thermocouple and the thermal bath. Consequently, a scaling factor must be applied to the recorded temperature oscillation, $\Delta T$. In this work, we determined this factor, $\kappa$, by enforcing thermodynamic consistency with the Ehrenfest relation. The Ehrenfest relation specifies how the jump in the ECE at a second-order phase transition is connected to the strain dependence of the ordering temperature, $\mathrm{d}T_c/\mathrm{d}\varepsilon$ \cite{Jerzembeck24,Ye22}. The latter quantity was directly determined in our measurements. Specifically, the Ehrenfest relation Eq.~\eqref{eq:Ehrenfest} (End Matter) can be rewritten \cite{Jerzembeck24} to

\begin{equation}
    \left( 1+\frac{C_n (T_c)}{\Delta C}\right) \frac{\Delta \eta}{\eta_n (T_c)}\,=\, -\left(1-\frac{1}{\eta_n(T_c)} \frac{\textnormal{d}T_c}{\textnormal{d}\varepsilon}\right),
    \label{eq:Ehrenfest2}
\end{equation}
with $C_n(T_c)$ ($\eta_n(T_c)$) being the heat capacity (the ECE) when approaching the transition temperature, $T_c$, from high temperatures and $\Delta C$ ($\Delta \eta$) the jump in the heat capacity (the ECE) at $T_c$. Since the ratio of ECE values $\frac{\Delta \eta}{\eta_n (T_c)}$ must be identical to the ratio of measured temperature oscillation amplitudes $\frac{\Delta (\Delta T)}{\Delta T_n (T_c)}:=\tilde{T}$, the value of $\eta_n (T_c)$ can be solely derived from experimentally-determined quantities through Eq.~\eqref{eq:Ehrenfest2}. The scaling factor $\kappa\,=\,\eta_n (T_c)/\Delta T_n (T_c)\,=130000(1000)$ was obtained for the lowest strain and at $0\,\mathrm{T}$, and was subsequently applied to all data sets. The above scaling approach is especially suited, since the correct magnitude of the ECE can be determined without explicit knowledge of the size of the applied AC strain $\varepsilon_\textrm{AC}$.

\subsubsection{Elastic properties of MnF$_2$}

The elastic tensor of MnF$_2$ were measured at room temperature in several earlier works. Taking the averages of Ref.~\cite{Har72} and references therein, we obtain (in GPa)

\begin{equation}
\mathbf{C} =
 \begin{pmatrix}
  101.9 & 79.7 & 70.9 &   0    &   0    &   0    \\
  79.7 & 101.9 & 70.9 &   0    &   0    &   0    \\
  70.9 & 70.9 & 165.3 &   0    &   0    &   0    \\
    0    &   0    &   0    & 30.9 &   0    &   0    \\
    0    &   0    &   0    &   0    & 30.9 &   0    \\
    0    &   0    &   0    &   0    &   0    & 69.7
\end{pmatrix}.   
\end{equation}

After rotating the matrix in-plane by 45$^\circ$, the in-plane Poisson's ratio can be calculated, $\nu\,=\,-\varepsilon_{1\,\bar{1}\,0}/\varepsilon_{1\,1\,0}=-0.07$.

\subsubsection{Detailed discussion of fit results for $T^*(\varepsilon,\mu_0H_z)$}

\label{sec:globalfit}

\textit{Results of the global fit to the mean-field equation --}

\begin{figure}
\includegraphics[width=0.48\textwidth]{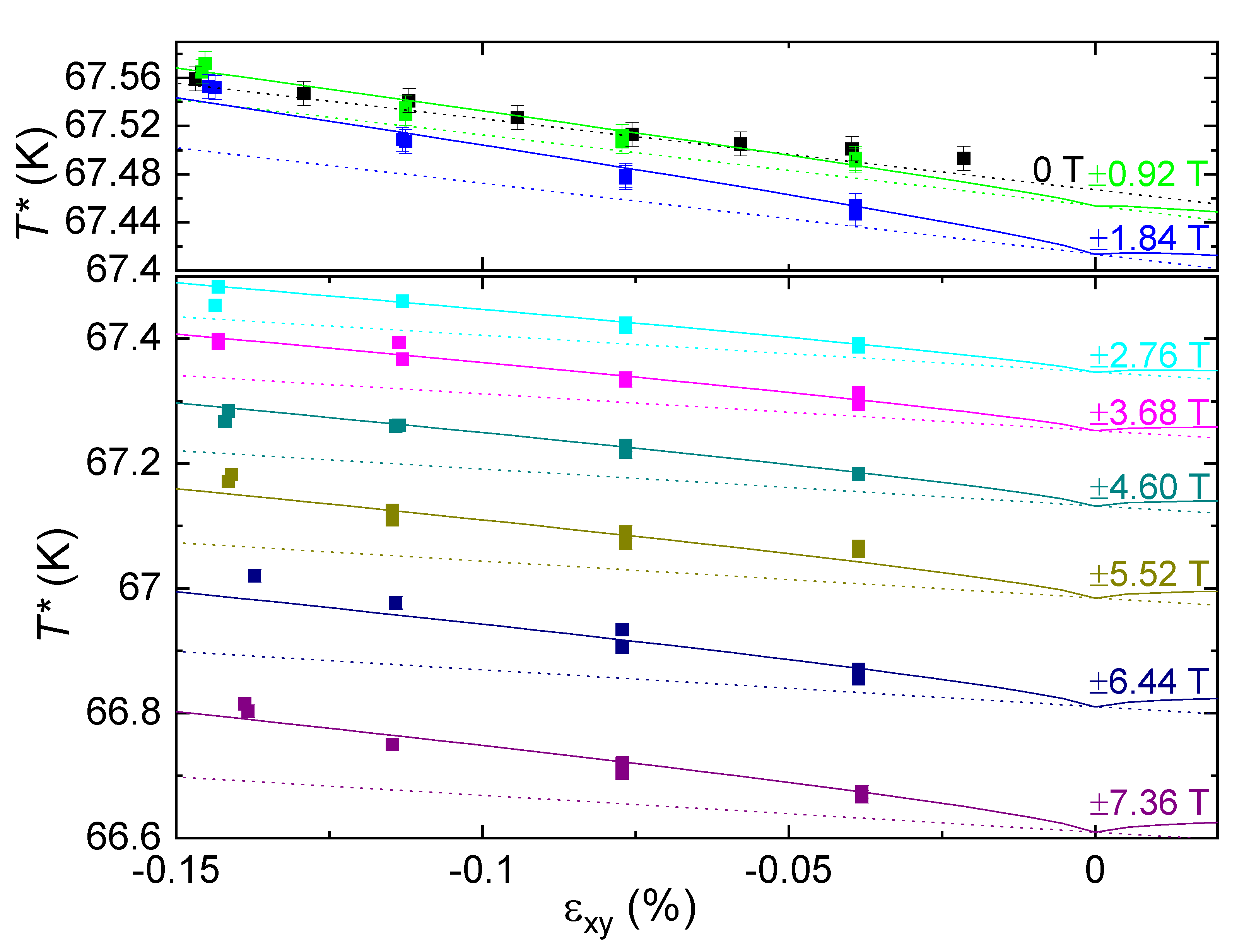}
\caption{\label{fig:TvsEpsdiffH}The transition temperature, $T_c$, at zero field and the crossover temperature, $T^*$, in finite fields, $\mu_0 H_z$, of MnF$_2$ as a function of applied strain, $\varepsilon_{xy}$, extracted from the elastocaloric data, shown in Figs.~\ref{fig:eps_data} and \ref{fig:H_data}. Dotted lines describe the response of the characteristic temperature which is not attributed to the AM response, i.e., $T_c$ (see Eq.~\eqref{eq:crossover-expanded} of the main text). The solid lines represent $T^*$ with the additional AM contribution $\lambda\,=\,0.56$. At 0\,T, these two models are identical and we chose to only show one of the model curves. Note that the data in the different panels is shown on differently scaled ordinates for better visibility. Data for magnetic fields above 1.84 T are omitted in the upper panel, while data for magnetic fields below 2.76 T are omitted in the lower panel to improve visibility.
}
\end{figure}

\begin{figure}
\includegraphics[width=\columnwidth]{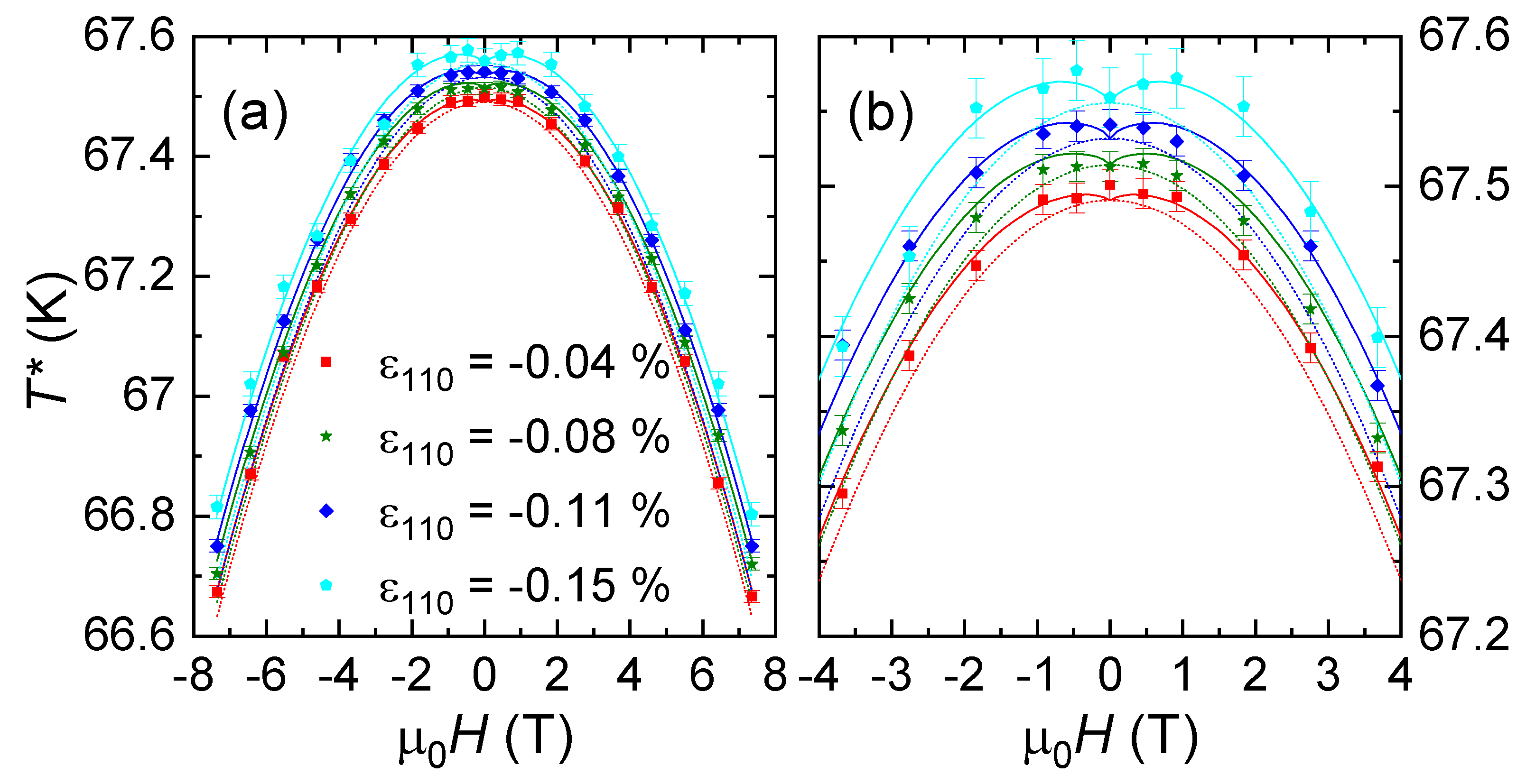}
\caption{\label{fig:TnvsH} The transition temperature, $T_c$, at zero field and the crossover temperature, $T^*$, in finite fields, $\mu_0 H_z$, of MnF$_2$ as a function of applied strain, $\varepsilon_{xy}$, extracted from the elastocaloric data, shown in Figs.~\ref{fig:eps_data} and \ref{fig:H_data}. Dotted lines describe the response of the characteristic temperature which is not attributed to the AM response, i.e., $T_c$ (see Eq.~\eqref{eq:crossover-expanded} of the main text). The solid lines represent $T^*$ with the additional AM contribution $\lambda\,=\,0.56$. (a) shows the full field range while (b) shows data up to about 4 T to improve visibility of the low field range.}
\end{figure}

In the following, we discuss details of the global fit of the $T^*(\varepsilon_{xy},\mu_0H_z)$ data to Eq.~\eqref{eq:crossover-expanded} of the main text. The obtained fit parameters are $\lambda\,=\,0.56(5)$, $T_c^{(0)}\,=\,67.467(3)$\,K, $a_1\,=\,59(3)$\,K, $a_2\,=\,0.01585(14)$\,K/T$^2$, with the error bars reflecting the fit error bars only. In Fig.~\ref{fig:TvsEpsdiffH} we plot the experimentally determined $T^*$ as a function of $\varepsilon_{xy}$ at different values of $\mu_0 H_z$ together with the result of the global fit as solid lines. The fit provides a good description of the experimental data over the full strain and field range. In this figure, we also include the expected behavior of the characteristic temperature, if $\lambda$ was zero, as dotted lines. This description without the AM contribution clearly deviates from the experimental data essentially for all finite strains and fields. 
In Fig.~\ref{fig:TnvsH} we show the same data as in Fig.~\ref{fig:TvsEpsdiffH} but as a function of $\mu_0H_z$ and at different values of $\varepsilon_{xy}$. This depiction stresses that the experimental data exhibits a kink at zero magnetic field, that can be attributed to the altermagnetic coupling to the conjugated field $\hat{h}$.

\begin{figure}
\includegraphics[width=0.48\textwidth]{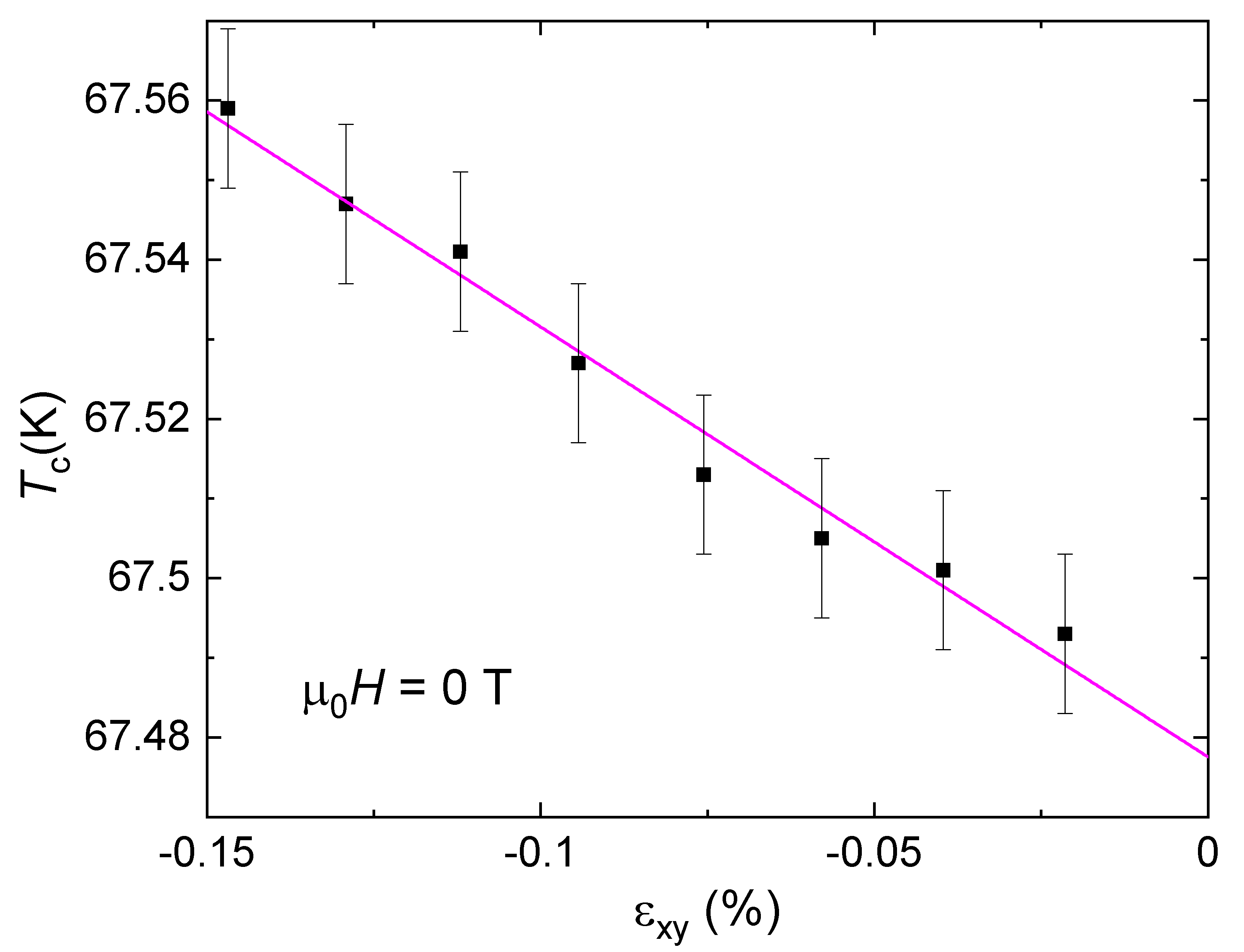}
\caption{\label{fig:TvsEps} The ordering temperature, $T_c$, of MnF$_2$ as a function of strain, $\varepsilon_{xy}$, extracted from elastocaloric measurements in zero magnetic field ($\mu_0 H\,=\,0\,$T). The data follows a linear behavior with a slope of $a_1\,=\,-54(3)\,\mathrm{K}$.}
\end{figure}

To demonstrate the reasonable choice of $a_1$ and $a_2$ of this global fit, we present next our experimental determination of the parameters $a_1$ and $a_2$, as defined by the free energy in Eq.~\eqref{eq:free-energy-expanded} of the main text. The parameter $a_1$ arises due to the symmetry-allowed coupling of $\Phi^2$ to symmetry-conserving  strains and results in an in-strain linear change of the ordering temperature $T_c$ with $\varepsilon_{xy}$. The transition temperatures, extracted from the data in Fig.~\ref{fig:eps_data} (End Matter), are shown in Fig.~\ref{fig:TvsEps}. The data are well described by a linear fit with slope of $a_1=-54(3)\,\mathrm{K}$, which agrees with the value of the global fit within the error bars.

\begin{figure}
\includegraphics[width=0.48\textwidth]{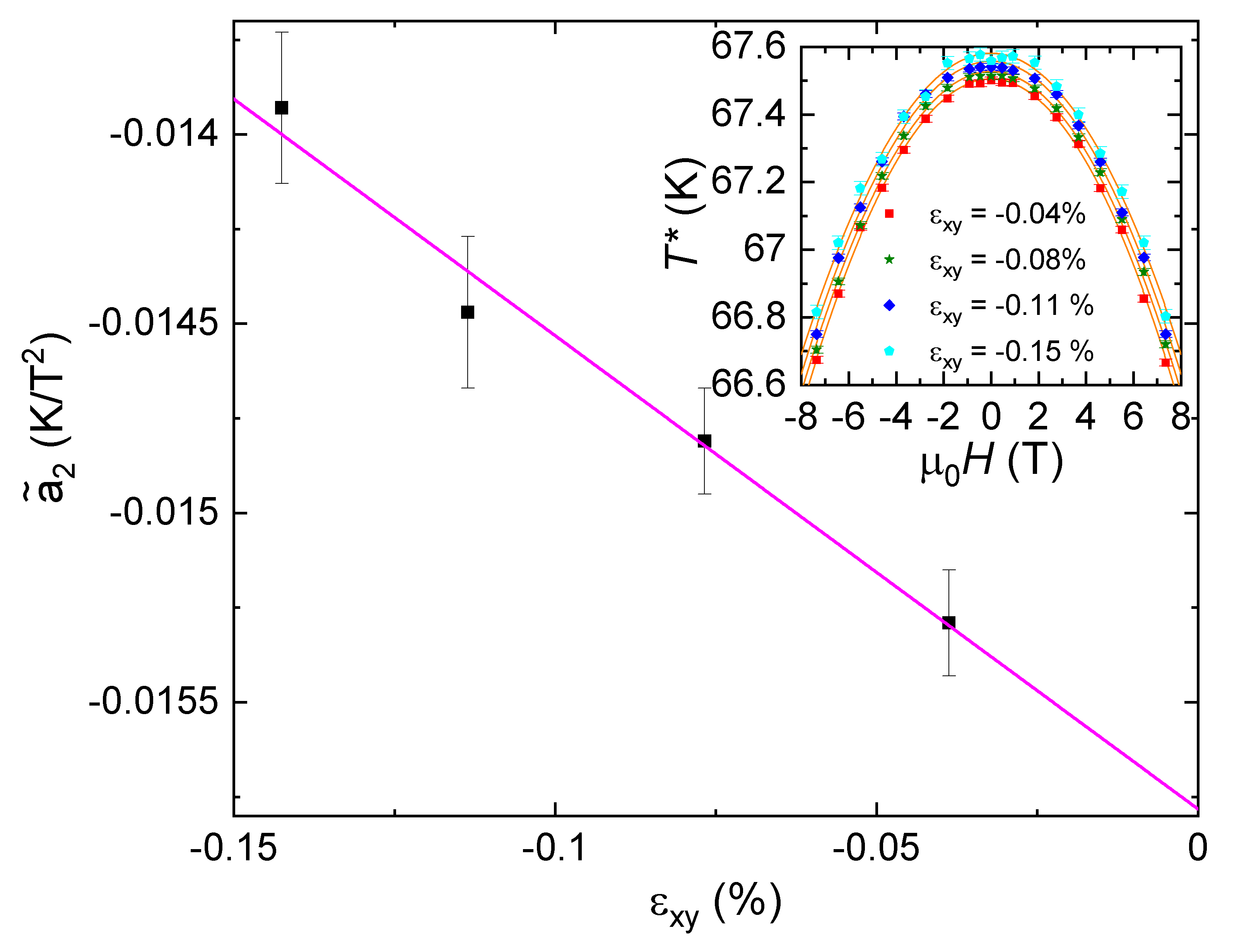}
\caption{\label{fig:bvsEps} Results of quadratic fits to the field dependence of the characteristic temperature at fixed strain, $\varepsilon_{xy}$. 
The quadratic coefficient, $\tilde{a_2}$, changes with applied $\varepsilon_{xy}$, as expected for a system with finite AM coupling, $\lambda$. A linear extrapolation to zero strain yields $a_2=-0.01578(9)$. The inset shows the field dependence of the characteristic temperature at fixed strain and the quadratic fits. 
}
\end{figure}

Next, we focus on the determination of $a_2$ at zero strain, i.e., the in-field quadratic suppression of the transition temperature. Since measurements at true zero applied stress (i.e., zero strain) are not possible in our experimental configuration (see Experimental Methods, Sec.\,\ref{sec:exp-methods}), we chose following procedure to determine $a_2$ independently from the global fit. For each strain, we fitted the $T_c^{*}$ vs. $\mu_0 H_z$ data (shown in the inset of Fig.~\ref{fig:bvsEps}) by a quadratic function 
and extracted the parameter $\tilde{a_2}$ (the quadratic coefficient of the fits). As shown in Fig.~\ref{fig:bvsEps}, $\tilde{a_2}$ decreases linearly in magnitude as higher the compression is. This change can be explained by the effect of a small $\lambda$, which becomes larger the higher the combined $\varepsilon_{xy}\mu_0H_z$ is (Note that the effect of $\lambda$ is still visible at small magnetic fields, where these quadratic fits provide an insufficient description of the experimental data, see the following section for a more detailed discussion). To determine $a_2$, $\tilde{a_2}$ can be extrapolated to zero strain. We obtain $a_2=-0.01578(9)$. This value agrees well with the one obtained from the global fit.

\textit{Discussion of symmetry-allowed higher-order terms --}

For completeness, we examine the influence of additional symmetry-allowed terms in the free energy, which could lead to modified functional dependencies of the characteristic temperature on strain and magnetic field. In particular, we consider the impact of potential higher-order symmetry-allowed contributions that would become relevant when both strain and magnetic field are present.

\begin{figure}[h!]
\includegraphics[width=0.48\textwidth]{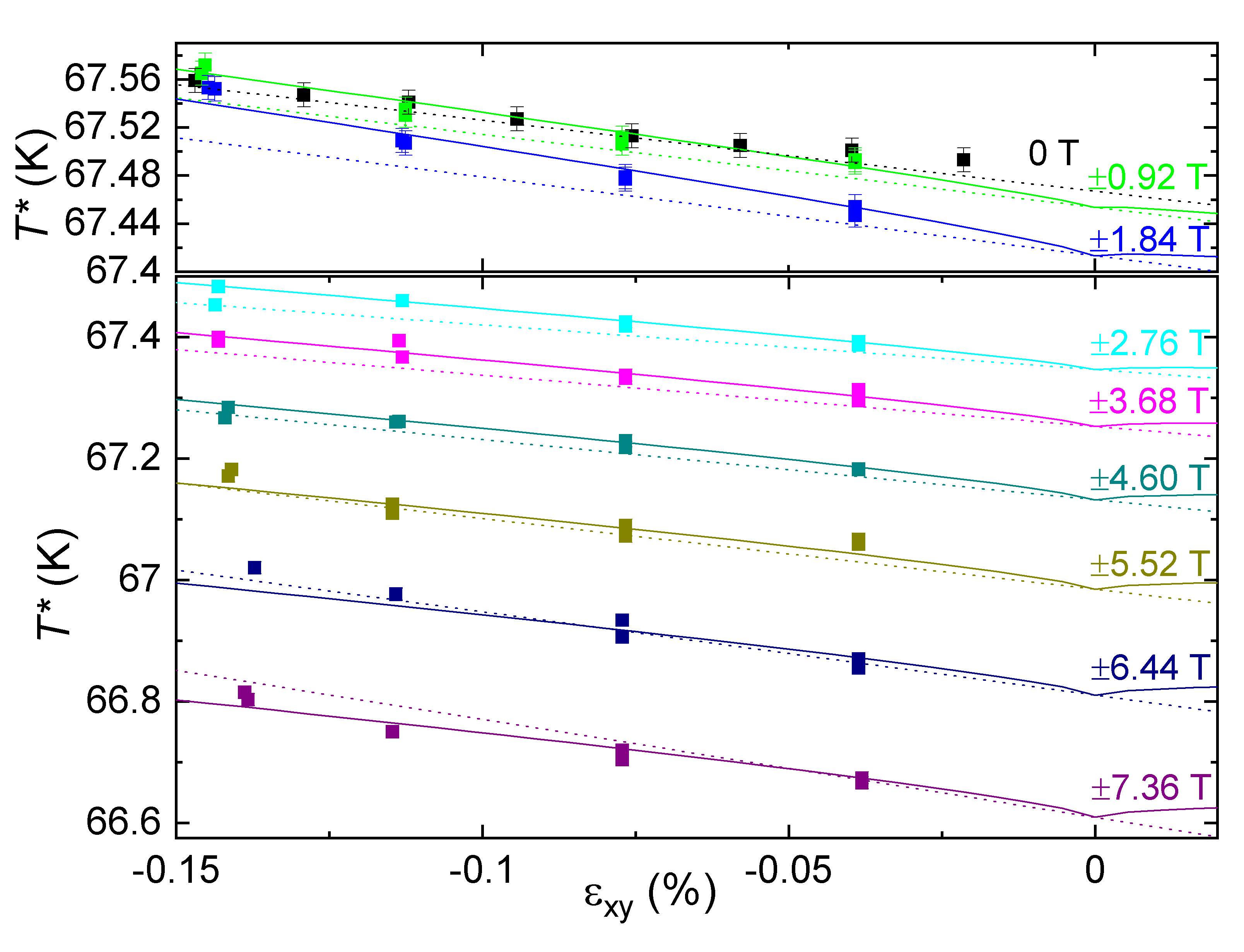}
\caption{\label{fig:TvsEpsdiffH2}The transition temperature, $T_c$, at zero field and the crossover temperature, $T^*$, in finite fields of MnF$_2$ as a function of applied strain, $\varepsilon_{xy}$, extracted from the elastocaloric data, shown in Figs.~\ref{fig:eps_data} and \ref{fig:H_data} (End Matter). The solid lines represent $T^*$ (see Eq.~\eqref{eq:crossover-expanded} of the main text) including the AM contribution (same as in Fig.~\ref{fig:TvsEpsdiffH}).  Dashed lines are global fits to $T^\prime_c\,=\,T_c^{(0)}+a_1 \varepsilon_{xy} + a_2^0(1+\gamma \varepsilon_{xy}) (\mu_0 H_z)^2$ (see Eq.~\eqref{eq:crossover-expanded} of the main text for comparison), i.e., with a strain-dependent quadratic contribution and no $\lambda$. Note that the data in the different panels is shown on differently scaled ordinates for better visibility. Data for magnetic fields above 1.84 T are omitted in the upper panel, while data for magnetic fields below 2.76 T are omitted in the lower panel to improve visibility.}
\end{figure}

In general, the next higher-order symmetry-allowed term in Eq.~\eqref{eq:free-energy-expanded} of the main text takes the form $\varepsilon_{xy} (\mu_0 H_z)^2 \Phi^2$. This is permitted because $\sigma_{110}$ also induces the symmetric strain component. Such a term effectively renders the coefficient $a_2$ strain dependent, i.e., $a_2 = a_2^0 (1+\gamma \varepsilon_{xy})$ (or equivalently the coefficient $a_1$ field dependent). If $\lambda$ were zero and $\gamma$ positive (while $a_2^0$ remains negative and $\gamma\varepsilon_{xy}<1$), this would lead to a positive $T^* - T_c$ under compressive strain for both positive and negative fields. However, in that scenario the $T^* - T_c$ data would not collapse onto a single curve when plotted against $\varepsilon_{xy} \mu_0 H_z$, nor would it exhibit a cusp at $\varepsilon_{xy} \mu_0 H_z = 0$. The behavior shown in Fig.~\ref{fig:1}\,(c) of the main text therefore strongly indicates that a model based solely on a strain-dependent $a_2$ without a finite $\lambda$ cannot account for our observations. Nevertheless, for completeness we compare  a model fit of $T^*$ with $\lambda$ finite and $\gamma\,=\,0$ to a model fit of $T_c$ with strain-dependent $a_2$, i. e., $\gamma\,=\,118(3)$, and $\lambda\,=\,0$ (but $T_c^{(0)}\,=\,67.467\,\mathrm{K}$, $a_1\,=\,-59\,\mathrm{K}$ and $a_2^0=-0.01585\,\mathrm{K/T^2}$ the same as before) in Fig.~\ref{fig:TvsEpsdiffH2}. The $T_c$ fit clearly does not describe the data at low fields and high strain values and provides an overall worse agreement with the experimental data than the fit to $T^*$.

Fitting our experimental data with both finite $\lambda$ and $\gamma$, while keeping the same $a_1$ and $a_2$ values as in the main text, yields $\lambda\,=\,0.48(3)$ and $\gamma =\,11(5)$. Because $\lambda$ is only slightly smaller than the value reported in the main text, we conclude that the behavior of $T^*(\varepsilon_{xy}, \mu_0 H_z)$ is predominantly governed by $\lambda$.

We note that a term of the form $\varepsilon_{xy}\,\cdot\,\mu_0H_z \Phi^2$ is not allowed by symmetry, if $\Phi$ is a $B_{2g}$ symmetric order parameter. Even if allowed, this term would not explain our observations, since it would result in an in-field asymmetric phase diagram, contrary to the results discussed in this Supplementary Information.

In the previous work by Ye \textit{et al.} \cite{Ye24} on an octupolar $f$-electron system, an additional higher-order  coupling term of the form $H_x H_y H_z \Phi$ was discussed. This term is not relevant in the present experimental configuration, since $H_x\,=\,H_y\,=\,0$.

\subsubsection{Comparison of ECE results under $\epsilon_{xy}\mu_0 H_z$ vs. $\epsilon_{xy}\mu_0 H_{110}$}

Based on symmetry arguments, a bilinear coupling of $\Phi$ is only allowed with $\epsilon_{xy}$ strain and $\mu_0 H_z$, but not with, e. g.,  $\epsilon_{xy}$ strain and $\mu_0 H_{110}$. Correspondingly, in order to demonstrate the symmetry-selectivity of our ECE data, we compare here the results of ECE measurements under $\epsilon_{xy} \mu_0 H_{110}$ with those under $\epsilon_{xy} \mu_0 H_{z}$, presented in the main text. For the $\epsilon_{xy} \mu_0 H_{110}$ configuration, we do not expect a change in the magnitude of the ECE near the critical temperature, nor do we expect the characteristic increase in the crossover temperature with applied field and strain. 

\begin{figure}
\includegraphics[width=0.48\textwidth]{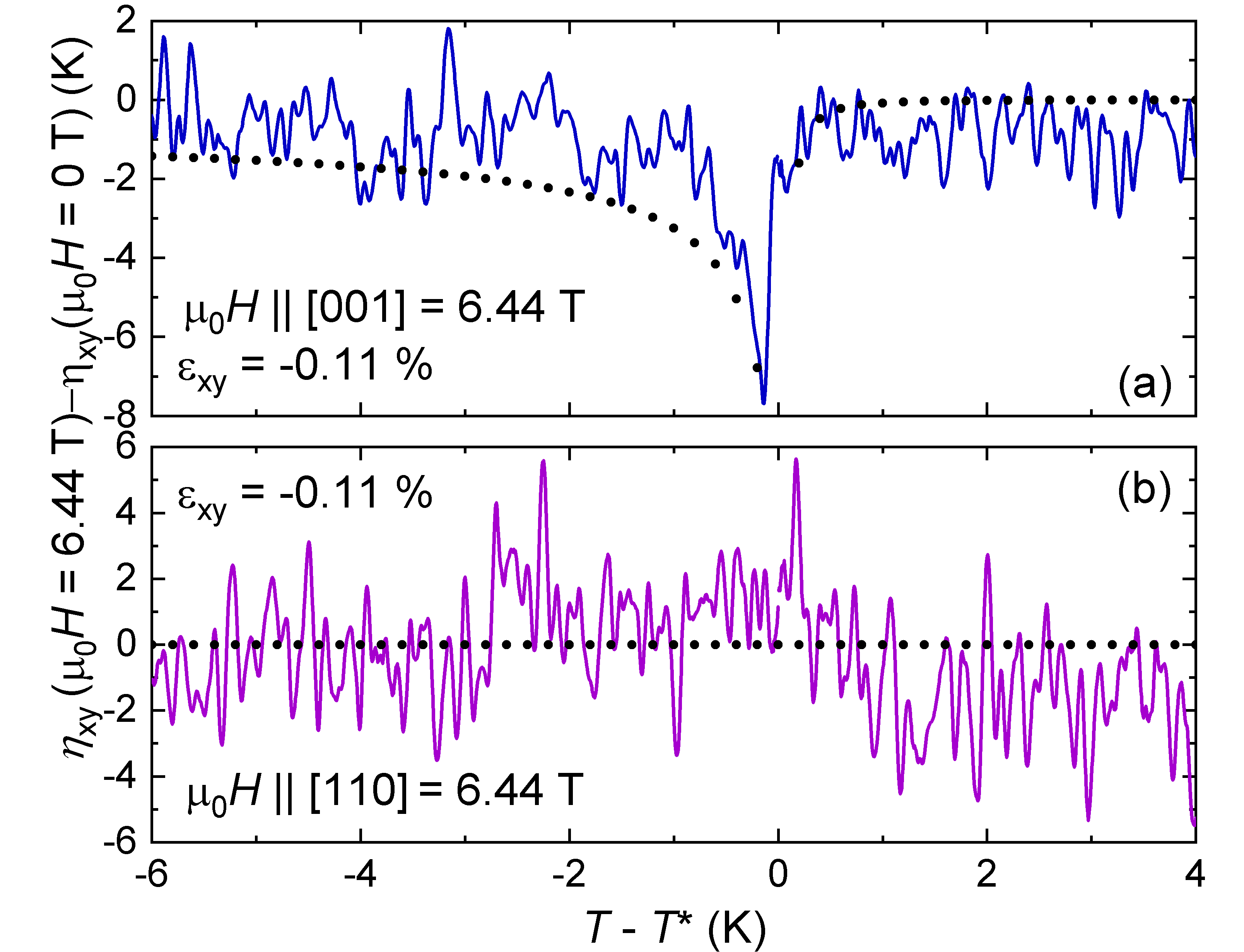}
\caption{\label{fig:fig5} The difference between the ECE at finite magnetic field and at zero magnetic field, i.e., $\eta_{xy} (\mu_0H=6.44\,\mathrm{T})-\eta_{xy} (\mu_0H=0)$ vs. a reduced temperature axis $T-T^*$. The upper panel shows experimental data (blue) and simulations (black) for $H_z\perp\epsilon_{xy}$. In the lower panel experimental data (purple) and simulations (black) for $H_{110}\parallel\epsilon_{xy}$ are shown.}
\end{figure}

First, to highlight that there is no change in the magnitude of the ECE near the critical temperature, we compare in Fig. \ref{fig:fig5} the difference of the ECE at a finite magnetic field to the ECE at zero magnetic field, i. e., $\eta_{xy} (\mu_0H\,=\,6.44\,\mathrm{T})-\eta_{xy} (\mu_0H\,=\,0)$, at the same finite strain and magnetic field for the two experimental configurations $H_z\perp\epsilon_{xy}$ (a) and $H_{110}\parallel\epsilon_{xy}$ (b). For this comparison, the data was plotted on a reduced temperature axis, $T-T^*$. In (a), we first plot the data from the main text in this specific representation. For temperatures lower, but close to $T^*$, there is a clear additional contribution to the experimental ECE in finite field. This feature also occurs in the simulations (dotted line) confirming its origin to be the AM coupling to strain and magnetic field in this symmetry-allowed configuration (see main text, Fig.\,\ref{fig:3}). In (b) we show data from measurements with $H_{110}\parallel\epsilon_{xy}$. Within the noise, there is no change in the magnitude of the ECE near the critical temperature, as expected from the absence of a bilinear coupling. Second, the extracted characteristic temperatures (data not shown) can be described by the equation for $T_c$ (Eq.~\eqref{eq:crossover-expanded}) solely, i.e., without a contribution of $\lambda$.

\end{document}